\documentclass[twocolumn,superscriptaddress,prx]{revtex4-2}
\usepackage[utf8]{inputenc}
\usepackage{graphicx}
\usepackage{tikz}
\usetikzlibrary{quantikz2}

\usepackage{hyperref, soul}
\usepackage[capitalize]{cleveref}

\hypersetup{
    colorlinks,
    linkcolor={blue},
    citecolor={blue},
    urlcolor={blue}
}

\usepackage{enumitem}

\newlist{tightitemize}{itemize}{1}
\setlist[tightitemize]{
  label       = \textbullet,
  labelindent = 0pt,      
  leftmargin  = 1em,      
  labelwidth  = 0.6em,
  labelsep    = 0.4em,
  topsep      = 2pt,      
  partopsep   = 0pt,
  parsep      = 0pt,
  itemsep     = 1pt,      
  align       = left
}

\usepackage{array}

\newlength{\rowgap}
\begin{document}

\title{
Error-Corrected Memory and Logic
on a Heavy-Hex Superconducting-Qubit Processor}

\author{Campbell K. McLauchlan}
\affiliation{School of Physics, The University of Sydney, Sydney, New South Wales 2006, Australia}

\author{Georgia M. Nixon}
\affiliation{School of Physics, The University of Sydney, Sydney, New South Wales 2006, Australia}

\author{Julien M. Drouet}
\affiliation{School of Physics, The University of Sydney, Sydney, New South Wales 2006, Australia}
\affiliation{School of Computer and Communication Science, Ecole Polytechnique Fédérale de Lausanne, Lausanne 1015, Switzerland}

\author{Xanda C. Kolesnikow}
\affiliation{School of Physics, The University of Sydney, Sydney, New South Wales 2006, Australia}

\author{Seok-Hyung~Lee}
\affiliation{Department of Quantum Information Engineering, Sungkyunkwan University, Suwon 16419, Republic of Korea}

\author{James Raftery}
\affiliation{IBM Quantum, T. J. Watson Research Center, Yorktown Heights, New York 10598, USA}

\author{Karthik Siva}
\affiliation{IBM Quantum, T. J. Watson Research Center, Yorktown Heights, New York 10598, USA}

\author{Stephen D. Bartlett}
\email{stephen.bartlett@sydney.edu.au}
\affiliation{School of Physics, The University of Sydney, Sydney, New South Wales 2006, Australia}

\author{Benjamin J. Brown}
\affiliation{IBM Quantum, T. J. Watson Research Center, Yorktown Heights, New York 10598, USA}
\affiliation{IBM Denmark, Sundkrogsgade 11, 2100 Copenhagen, Denmark}

\author{Robin Harper}
\affiliation{School of Physics, The University of Sydney, Sydney, New South Wales 2006, Australia}

\begin{abstract}

Large quantum computations require high-fidelity error-corrected memory and fault-tolerant logical operations. To achieve these with superconducting circuits, quantum processors will need many components operating with low error rates.
In practice, fabrication of monolithic superconducting devices produces unavoidable heterogeneity in component performance, which the quantum computer must accommodate. We implement strategies to operate a large dynamic compass code on an array of 156 superconducting qubits arranged on a heavy-hexagonal lattice, making use of the entire monolithic quantum processor and redesigning our measurement schedule to exclude underperforming components. We demonstrate improvements of up to $35\%$ in logical error probabilities for a memory experiment by using a noise-informed decoder combined with strategies to exclude the worst-performing qubits and couplers, along with post-selecting on shots with a low probability of leakage errors. In addition to these memory results, we also observe performance gains of up to $56\%$ when these strategies are applied to measurement-based logic, demonstrated by stability experiments. We furthermore demonstrate high-fidelity fault-tolerant logic through lattice surgery. Through \textit{pre}-selection filtering techniques, we can obtain Bell fidelities above $77\%$, and further logical gap post-selection using exclusive decoding increases these fidelities to above $97\%$. Our experiments demonstrate that high performance can be maintained in key elements of fault-tolerant quantum memory and logic when working on monolithic superconducting quantum processors with underperforming components, by using a suite of noise-informed quantum error correction strategies.
\end{abstract}

\maketitle

\section{Introduction}

Superconducting platforms offer scalable routes towards fault-tolerant quantum computing. They admit short error-correcting cycle times resulting from fast operations~\cite{google24}, high-yield device fabrication and advanced characterization techniques~\cite{van_damme_advanced_2024,lang2025characterization,Lang_CMOS_2025}, long coherence times~\cite{bland_millisecond_2025,tuokkola_methods_2025,jiang_advancements_2025}, high-fidelity two-qubit gates~\cite{marxer99fidelity2026,jiang_advancements_2025}, and leakage mitigation strategies~\cite{PRL_leakage2025,mcewen2021,McEwen2023,miao_overcoming_2023,Lacroix2025}. Key primitives for fault-tolerant quantum computing have been demonstrated~\cite{google24,google_magic2025,Ye_magicstateprep_2023,Gupta2024,wang2026super,drouet2026defects,harper2025characterising,lacroix2024scalinglogiccolorcode,caune_demonstrating_2024,besedin_lattice_2026,bence2024, lange2026}, and theoretical work has outlined pathways to achieving utility with superconducting quantum devices~\cite{Gidney2021,gidney2025factor2048bitrsa,yoder2025tourgrossmodularquantum}.

To operate quantum computers at scale, we must perform high-fidelity error-correction and fault-tolerant logic across large sections of devices. Superconducting platforms pose several challenges in this regard. First, one must design a code and architecture that can be operated in a scalable fashion given a fixed connectivity. Second, we must accurately decode errors, requiring us to obtain good noise-characterization data and measurement information in a timely and efficient manner. Third, any approach must allow for some number of defective or underperforming components. Without this, superconducting devices may face prohibitively low manufacturing yields, or the codes' performance will suffer when operating on large-scale chips~\cite{drouet2026defects}. 
Finally, scaling up to larger codes introduces more points at which qubits may leak out of the computational subspace, which can dramatically deteriorate performance~\cite{Brown_cross_brown_2020}.

In this work we demonstrate the operation of key elements of error-correction and logic across the entirety of a 156-qubit superconducting quantum computing device with a heavy-hex connectivity, \texttt{ibm\_boston}, finding substantial performance gains by using noise-informed approaches. We benchmark our error-correction performance through quantum memory experiments, in which logical qubits in a variety of stabilizer states are prepared, and then measured after several rounds of syndrome extraction. Memory experiments help determine the ability of our error-correcting scheme to protect quantum information while it is idling in a processor. We then benchmark fault-tolerant logic performance through stability experiments~\cite{Gidney2022dual,harper2025characterising,zen2026low}, which characterize the ability to perform logical measurements and manipulate codes, critical for operations such as lattice surgery~\cite{Horsman2012,fowler2012b,Litinski2019gameofsurfacecodes,fowler2019lowoverheadquantumcomputation} and logical Clifford gates~\cite{Litinski2019gameofsurfacecodes,Geher_2024,Gidney_2024_ybasis}. 
Finally, we perform lattice surgery experiments directly~\cite{Bombin_2009, fowler2012b, Horsman2012,Gupta2024,RyanAnderson2024,lacroix2024scalinglogiccolorcode,besedin_lattice_2026, Raussendorf2007, Brown2017, Litinski2019gameofsurfacecodes, Cohen2022,Thomsen2022,Davydova2025,bodeker2026} to measure the performance of fault-tolerant logic in the device. These are sensitive to logical errors in both memory and stability. 
These logical characterization experiments help determine the capability of the approaches and devices for fault-tolerant quantum computing.

Our code of choice is the dynamic compass code~\cite{lee2026scalable, zen2026low}. It is designed for the heavy-hex qubit array~\cite{Chamberland2020topological} and has a relatively low footprint for a given code distance, in comparison to other codes one might consider on this qubit geometry~\cite{McEwen2023,Benito_2025,vezvaee2025surfacecodescalingheavyhex,Hastings2021dynamically,Gidney2022,Kesselring2024,Davydova2023floquet,Setiawan2025}. The dynamic compass code also demonstrates a threshold in both memory and stability~\cite{zen2026low}, meaning theoretically that it can be scaled to improve performance in memory and logic.

We demonstrate the benefits of noise-informed decoding in memory experiments operated over the entire device, as well as the utility of this approach in decoding stability and lattice surgery experiments. We use averaged circuit eigenvalue sampling (ACES)~\cite{flammia2021b,hockings2024,hockings2025improvingerrorsuppressionnoiseaware} to extract circuit-specific Pauli noise models that inform decoding. 
We can obtain more information by extracting data from measurement pulses beyond the binary outcomes, and using this data for soft decoding
~\cite{pattison2021improvedquantumerrorcorrection,Ali2024,Majaniemi_2026}. We show that soft information and ACES noise data interact in stability performance in ways not previously observed. In particular, soft information improves the decoding of stability experiments only when ACES information is not used, bringing performance with a decoder informed by a simpler noise model on par with ACES-informed decoding.

As larger superconducting chips are manufactured, a degree of heterogeneity in the performance of components such as couplers and qubits, and even completely defective components, will need to be tolerated in order to maintain acceptable yields from the manufacturing processes. The quantum processor \texttt{ibm\_boston}, for example, typically displays very low median error rates, but has a small number of qubits or couplers suffering from noise $1$ -- $2$ orders of magnitude higher~\cite{ibm_compute}. We therefore seek quantum error correction strategies that can maintain high performance despite this heterogeneity. 
There has been considerable recent effort towards designing defect-exclusion strategies for the surface code~\cite{Stace2009thresholds, Auger2017, Strikis2021, Siegel2023adaptivesurfacecode, Lin2024codesign, Wei2025low, Leroux2025snakes, Debroy2025luciinsurfacecode, mishmash2025excising,  wolanski2026automated, wei2026adaptivedeformationcolorcode,higgott2025handlingfabricationdefectshexgrid,anker2025optimizedmeasurementschedulessurface} followed by some demonstrations in experiment~\cite{drouet2026defects,kim2026luciibmhardwareerror}. However, comparatively little work has been devoted to similar strategies for codes on layouts with more general connectivities, and in particular, lattices that are less connected than a square grid, making defect avoidance more challenging~\cite{McLauchlan2024accommodating,higgott2025handlingfabricationdefectshexgrid,wolanski2026automated}. Here, we design and implement an approach for defect-mitigation in the dynamic compass code operated on the heavy-hex qubit array. Our strategies improve performance substantially in both memory and stability in \texttt{ibm\_boston}, beyond what is achievable with noise-informed decoding alone.

Measurement data also provides information about when leakage has occurred~\cite{lee2026scalable,drouet2026defects,sundaresan2022,Krinner2022}. We can obtain estimates for the role of leakage in limiting performance by post-selecting on runs in which it is likely that no qubits have leaked. For our lattice surgery results, we introduce a technique for leakage \textit{pre}-selection~\cite{barron2026observableestimationabsenceclassical}, in which we do not execute runs when the starting state is likely to have already experienced some leakage. We also investigate the effect of pre- and post-selection more broadly. We find substantial improvements from removing shots that prepared an initial code state with error, and also from using logical gap post-selection~\cite{Log_gap_2024,gidney_yoked_2025,meister2024efficientsoftoutputdecoderssurface,Smith2024}. Finally, we observe that post-selecting on the convergence of a belief propagation stage of the decoder~\cite{higgott2023improved} results in similar improvements in a different experiment.

\subsection{Summary of Results}
We here present a summary of our results, presented in detail in \cref{sec:Results}.
We introduce defect exclusion techniques for the dynamic compass code in Section~\ref{sec:dynamic_compass_code}. When combining  ACES-informed BeliefMatching decoding, defect mitigation and leakage post-selection, in full-device memory experiments, we observe $4.11\%$ and $2.56\%$ logical error probabilities per round in the $Z$ and $X$ bases, respectively. 
These are $32.3\%$ and $35.2\%$ improvements compared to baseline performance, in the two bases, respectively (Section~\ref{sec:memory_results}). These results were qualitatively consistent across multiple days of data collection. Our best results, without any post-selection, found $4.4\%$ and $2.9\%$ logical error per round in the two bases, respectively. On a different day in which the noisy components were performing worse, we observed improvements resulting just from defect exclusion of up to $40\%$, depending on the decoder used. We observe that defect exclusion still produces sizeable gains even with good noise characterization. 
We find that our $7\times 7$ memory patch with defect-mitigation meets or surpasses the average $5\times 5$ memory patch performance in both bases, indicating a key role for defect exclusion strategies in pushing monolithic devices into the sub-threshold regime. 

In stability, we observe continual error-suppression up to 30 rounds of syndrome extraction only when we exclude defects, whereas we find plateaus in patches without defect mitigation starting around 12 rounds. At 30 rounds, we find an overall improvement of $56\%$ in failure probability from our techniques (Section~\ref{sec:stability_results}). We observe that soft information improves performance only when using reported (not ACES) noise for decoding.

When performing lattice surgery, we obtain lower bounds on Bell fidelities of $77.3\%$ when just using pre-selection filters. When combining this with logical gap post-selection, we obtain a Bell fidelity above $97.6\%$ with a post-selection yield of $27\%$ (Section~\ref{sec:lattice_surgery}). In the Appendices, we report further results for memory, stability and lattice surgery (including logical state teleportation) experiments.

\begin{figure*}
\begin{tikzpicture}
\node at (0,0) {\includegraphics[width=0.95\textwidth]{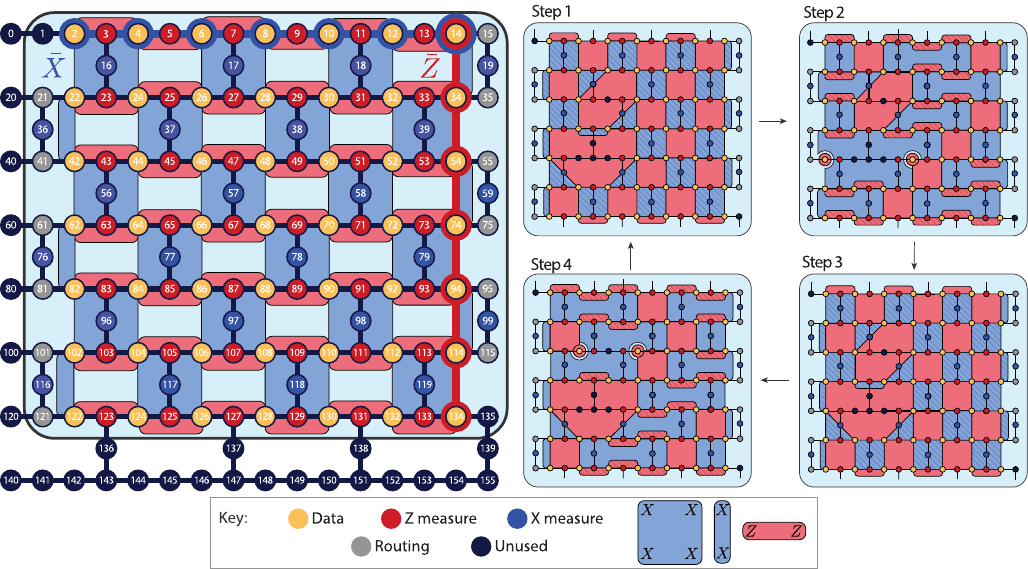}};
\node at (-8.5,4.8) {(a)};
\node at (0,4.8) {(b)};
\end{tikzpicture}
\caption{\textbf{Dynamic compass code on the 156-qubit quantum processor \texttt{ibm\_boston}.}  (a) Arrangement of qubits and couplers on \texttt{ibm\_boston}, with the layout of the $7\times 7$ dynamic compass code shown. Data qubits are shown in yellow, $Z$ and $X$ measure qubits in red and blue respectively, and qubits used only for routing and flag information in gray. The $Z$ ($X$) checks of the code are shown with red (blue) plaquettes. The $Z$ ($X$) logical operator of the code is shown with a red (blue) string and rings. The code arises from a four-step sequence of check operator measurements. (b) Syndrome extraction schedule of the dynamic compass code with underperforming qubits 85 and 46 excluded. In each step, shaded checks indicate parity measurements made during that step, while unshaded colored regions indicate members of the instantaneous stabilizer group.\label{fig:Main_code_patch}}
\end{figure*}

\section{Results}\label{sec:Results}

\begin{figure*}
\begin{tikzpicture}
    \node at (0,0) {\includegraphics[width=0.54\textwidth]{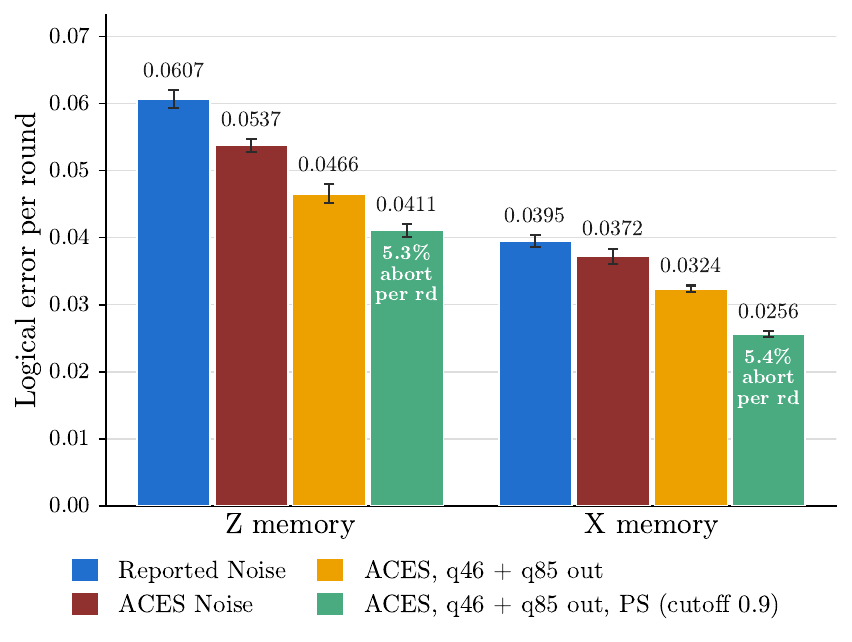}};
    \node at (9,0) {\includegraphics[width=0.43\textwidth]{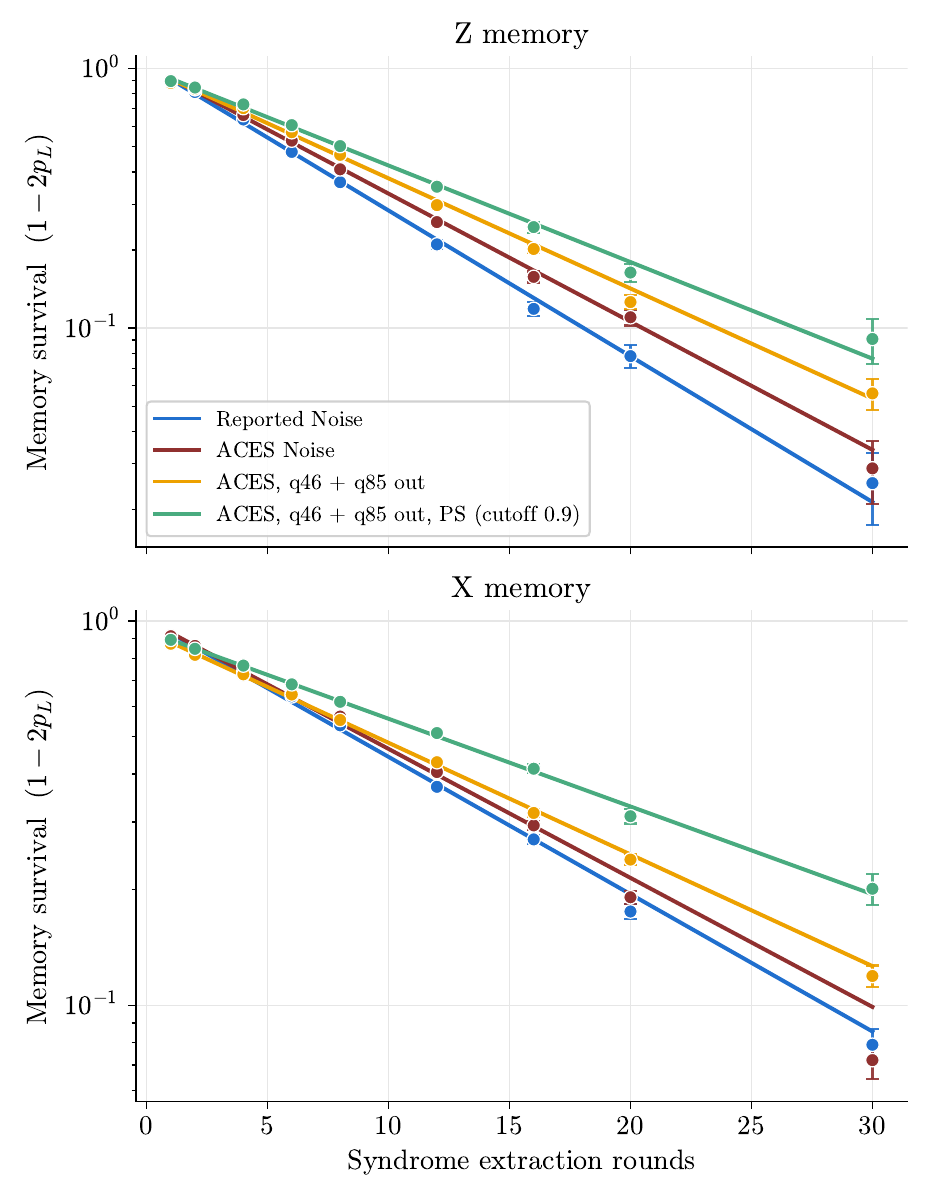}};
    \node at (-4,4) {(a)};
    \node at (4.8,4.0) {(b)};
\end{tikzpicture}
    \caption{\textbf{Memory experiments for $7\times 7$ dynamic compass code.} Results of memory experiments performed in both Pauli $Z$ and $X$ bases, comparing the improvements from decoder, code and post-selection modifications. (a) Logical errors per round for four different noise-informed strategies, reported in detail in Table~\ref{tab:noise_models}. First, we inform a BeliefMatching decoder with generic, IBM-reported noise information  (blue data). We use this information to construct a Pauli error model (see Section~\ref{app:reported_noise}) that sets the prior probabilities in the belief propagation step. We compare this to a decoder that is provided ACES noise information (red data). The improved logical error probabilities indicate ACES offers a more accurate characterization of the system during syndrome extraction. We then compare these results to a modified code in which we perform active defect exclusion on qubits 46 and 85, along with their surrounding couplers (orange data). 
    Finally, we compare the above to the result of excluding defects \textit{and} removing leakage from analyzed shots through post-selection (at a cutoff probability of $0.9$; light green data). All data was generated on 2026/06/01.  The logical error per round is obtained by fitting the measured logical error probability $p_L$ to an exponential function of the number of measurement rounds $r$ after the logical subspace is set up (from $r=$ 1 to 30), as explained in the main text.
    (b) Memory survival probability as a function of number of syndrome extraction rounds, for $X$ and $Z$ memory experiments and all decoders, defect exclusion and post-selection conditions mentioned above. \label{fig:decoder_defect_improvements}}
\end{figure*}

\subsection{Dynamic Compass Code for a Heavy-Hex Qubit Array Excluding Defective Components}\label{sec:dynamic_compass_code}

We perform experiments using the IBM Quantum \textit{Heron} class of superconducting quantum processor, where 156 qubits are arranged on a plane with connectivity described by a heavy-hex lattice, displayed in \cref{fig:Main_code_patch}.
We present results on the device \texttt{ibm\_boston}, which supports mid-circuit measurements and high fidelity (median $\simeq 99.7\%$) two-qubit gates. The fidelity of processor components vary day-to-day, but underperforming components tend to cluster around one valence-2 qubit at site $46$, and one valence-3 qubit at site $85$ (see Sec.~\ref{sec:defect_exclusion_details} for discussion of defective qubits and couplers in the device).

\begin{figure*}
\begin{tikzpicture}
\node at (0,0) {\includegraphics[width=0.95\textwidth]{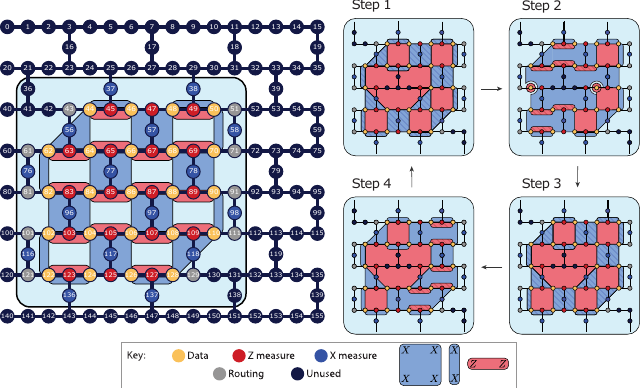}};
\node at (-8.5,5.0) {(a)};
\node at (0.5,5.0) {(b)};
\end{tikzpicture}
\caption{\textbf{Stability experiment for the dynamic compass code on \texttt{ibm\_boston}.} (a) We perform stability experiments using the $5\times 5$-sized patch shown in its location on the device.
The qubit color conventions are the same as those used in Fig.~\ref{fig:Main_code_patch}. 
The product of $X$ checks in any time step across the entire patch evaluates to $+1$ in the absence of noise. (b) We perform defect exclusion using the same methods as in memory experiments (see Section~\ref{sec:Methods}). Once excluding defective components around qubit 85 (excluded qubits shown in black), the period-4 measurement sequence is shown, with unshaded blue/red plaquettes indicating the $X$/$Z$ members of the instantaneous stabilizer group, respectively. Shaded plaquettes indicate checks that are measured in that step.\label{fig:stability_patch}}
\end{figure*}

We run 
experiments with the dynamic compass code~\cite{zen2026low,lee2026scalable}, and test strategies to treat the defective components. This code is compatible with the heavy-hex qubit array and demonstrates a threshold in both logical bases~\cite{zen2026low}. The code is a dynamic, or Floquet, code with $X$-checks ($Z$-checks) consisting of four-body (two-body) measurements on data qubits. The code is designed from the gauge measurements of the heavy-hex code~\cite{Chamberland2020topological,harper2025characterising} for use on the heavy-hex layout. We consider a $7\times 7$ code for the memory experiment as shown in Fig.~\ref{fig:Main_code_patch}, and a $5\times 5$ code for the stability experiment as shown in Fig.~\ref{fig:stability_patch}.
The $7\times 7$ memory patch is the largest $n\times n$ patch that fits on the device. Both patches are large enough that they cannot avoid operating across regions with defective components.

Our sequence for the $7\times 7$ code with defect exclusion is shown in Fig.~\ref{fig:Main_code_patch}(b). Importantly, the excluded qubits (black dots) and their neighboring couplers are never used. Instead, we redesign gauge measurements around the defective qubits to form superstabilizers that are not supported on the defective components. This results in some $X$ gauge checks being reduced to weight-$2$ or $3$, and some $Z$-gauge checks reduced to weight-$1$.

Additionally, the dynamic nature of the code allows for all stabilizers to remain low weight.
The dynamic compass code~\cite{zen2026low, lee2026scalable} does not maintain a fixed stabilizer group; we measure a sequence of anti-commuting checks that changes the instantaneous stabilizer group over a finite period~\cite{Hastings2021dynamically,alam2025baconshorboardgames}.
The four-step periodic sequence of check measurements is shown in  Fig.~\ref{fig:Main_code_patch}(b).  In Steps 1 and 3 we measure all $X$-checks in the code. In Step 2 we measure a subset of the $Z$-checks and in Step 4 we measure the remainder of the $Z$-checks. 
The partition of $Z$-checks into Steps 2 and 4 allows for $X$ stabilizers in the instantaneous stabilizer group of each step to remain low weight.  
Away from defects, they have a maximum weight of 8 on data qubits, while near defects this may increase up to 10.

The four-step sequence can be parallelized to only require two measurement layers~\cite{harper2025characterising}. Since $Z$ and $X$ check measurements occur on disjoint subsets of measure qubits, we can measure $X$ and $Z$ checks simultaneously. Therefore, we delay the measurement of $X$ measure qubits in Step 1 (Step 3) until the $Z$ measure qubits are ready to be measured in Step 2 (Step 4). 
Aligning the measurement layers reduces the overall time taken for syndrome extraction as well as data qubit idling, since measurements consume the majority of wall-clock time of a syndrome extraction cycle. 
See Section~\ref{sec:syndrome_mmt} for further details on the syndrome measurement circuits around defects.

\subsection{Memory Experiments}\label{sec:memory_results}

\begin{figure}
    \centering
    \includegraphics[width=0.99\linewidth]{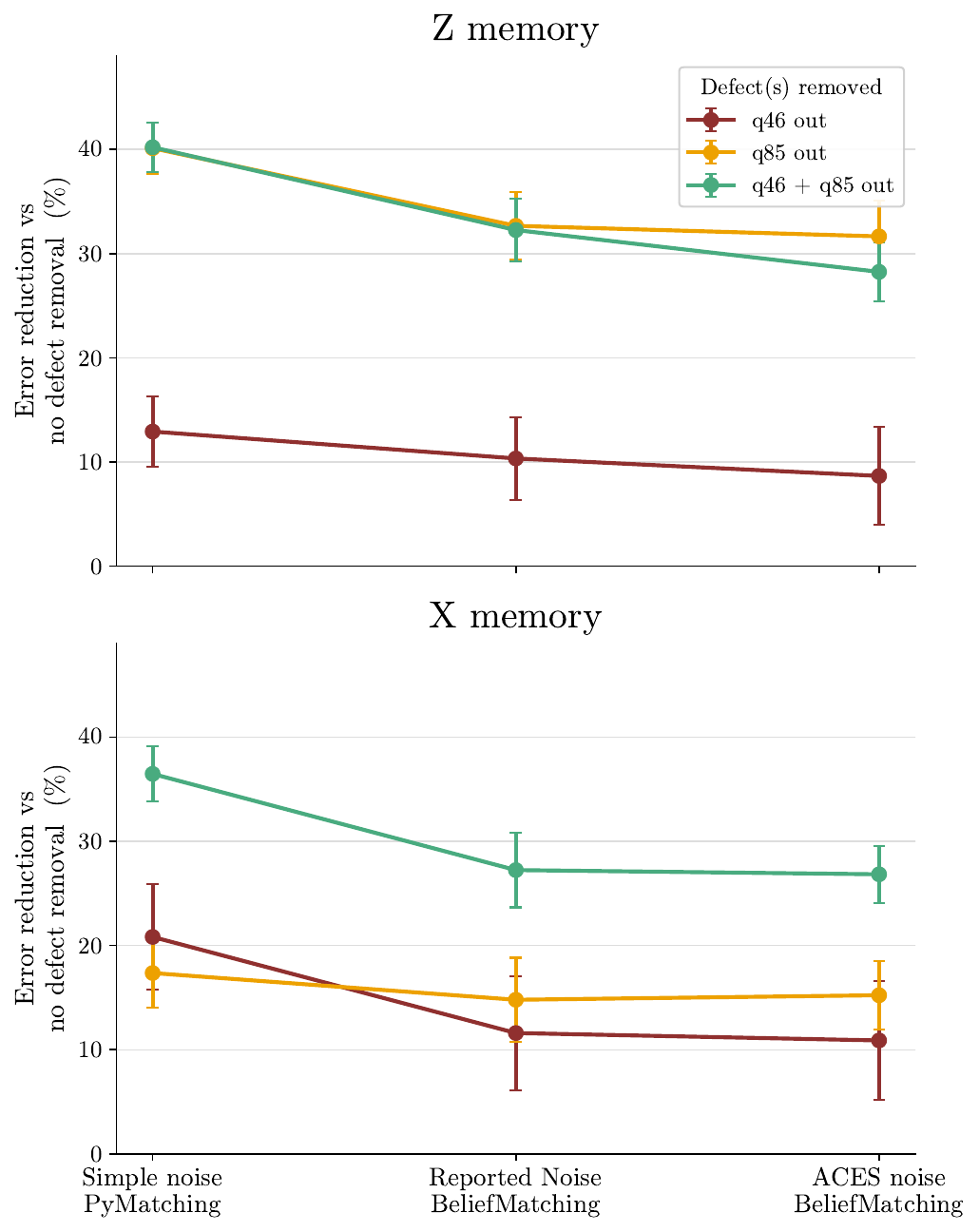}
    \caption{\textbf{Defect exclusion and decoding strategies.} We compare excluding no defects, excluding qubit 85, excluding qubit 46, or excluding both (this patch is shown in Fig.~\ref{fig:Main_code_patch}). We also compare three different decoders: PyMatching with a simple uniform noise model, BeliefMatching with a reported noise model, and BeliefMatching with an ACES-informed noise model. Even with better noise information and more accurate decoders, the gains of defect exclusion remain pronounced. We plot the percentage reduction in logical error per round compared to no defect-mitigation for the three defect exclusion choices and three decoders. Data was collected on 2026/05/02.\label{fig:dropout_vs_decoder}}
\end{figure}

Using the $7 \times 7$ dynamic compass code defined in the previous section, we perform memory experiments to characterize its performance. 
In these experiments, we initialize the code in a logical stabilizer state (either $\ket{0}$, $\ket{1}$, $\ket{+}$ or $\ket{-}$), perform a number of rounds of syndrome extraction, counted by the number of layers of measurements in the circuit
after initialization, and then measure the data qubits to determine the state of the logical qubit fault-tolerantly. We average over the two basis state preparations ($\ket{0}$ and $\ket{1}$, or $\ket{+}$ and $\ket{-}$) for each memory experiment ($Z$ or $X$).

Figure~\ref{fig:decoder_defect_improvements}(a) compares the logical error per round using the four noise informed decoding strategies presented in Table~\ref{tab:noise_models}.
In all cases, we use a BeliefMatching decoder~\cite{higgott2023improved} informed by noise characterization data. In the first case (reported noise), we use a Pauli noise model based on the coupler and qubit error information provided by IBM, specific to the day of the experiment (see Section~\ref{app:reported_noise}). In the second case (ACES noise), we generate noise models based on ACES noise characterization experiments run contemporaneous with the corresponding memory experiments. In some cases, ACES noise information obtained from a different day or that was specific to a slightly different syndrome extraction circuit needed to be used due to absence of more accurate data. 
The ACES noise model captures more of the device inhomogeneity than the reported noise, incorporating, for example, circuit-layer-specific differences in gate noise profiles resulting from other gates that are operated simultaneously. The more accurate noise model results in modest improvements over the reported noise data. 

Excluding underperforming components around qubits 85 and 46 resulted in further improvements. This defect exclusion consistently resulted in a reduced logical error per round, on every day of data collection. The combination of ACES information and defect exclusion improved the logical error per round by $23.2\%$ and $ 18.0\%$ for the $Z$ and $X$ basis memory experiments, respectively, compared to reported noise decoding.

We see continued improvements in logical error per round when recalibrating measurements using the IQ data and removing shots that may have experienced leakage through post-selection. In the $Z$ basis, we find an overall improvement in logical error of $32.3\%$, and in the $X$ basis we find an improvement of $35.2\%$.

The IQ plane data is obtained from measurements, which carry more than just binary classification information. 
To perform leakage post-selection, we first prepare and measure all qubits in states $\ket{0}$, $\ket{1}$ and $\ket{2}$, and fit the resulting IQ data spreads to weighted Gaussian mixture models (see Ref.~\onlinecite{lee2026scalable}). From these models we can use the IQ data from QEC experiments to determine the probability that each measurement detects a leakage event. We discard any shots from the analysis in which we find a probability of $\geq 90\%$ that the qubit has leaked. Furthermore, for the shots that are not discarded, we use the IQ data to produce probabilities $p_0$ and $p_1$ that the qubit is in state $\ket{0}$ or state $\ket{1}$, respectively, and take the measurement outcome to be $\text{argmax}_i\; (p_i)$. The use of IQ data and post-selection results in a further reduction in logical error per round of $11.8\%$ and $21.0\%$ in $Z$ and $X$ bases, respectively (see Fig.~\ref{fig:decoder_defect_improvements}).
For all data, we use a BeliefMatching decoder~\cite{higgott2023improved}. 

We note that to calculate logical error per round in Fig.~\ref{fig:decoder_defect_improvements}(a), we fit the logical error rate data, averaged over both basis states for that memory experiment, to an exponential decay $p_L = \frac{1}{2}(1-Ae^{-t/\tau})$ and extract the logical error per round from that fit, which is $\epsilon_L = \frac{1}{2}(1-e^{-1/\tau})$ where the uncertainties are propagated from the fitted decay constant $\tau$. Figure~\ref{fig:decoder_defect_improvements}(b) shows the memory survival $1 - 2 p_L$ per syndrome extraction round for both $X$ and $Z$ memory. For each data point we collect $4$ runs of $2048$ shots (runs have longer separations than shots), all on the same day, and aggregate together all runs when calculating the logical error probabilities.

We ultimately find a logical error per round of $2.56\% - 4.11\%$ when operating this full-device memory experiment (performing leakage post-selection). Even without leakage post-selection, our best memory results achieve similar logical errors per round (see Appendix~\ref{app:further_mem_data}). The best logical errors per round observed for memory were $4.4\%$ and $2.9\%$ in the $Z$ and $X$ bases, respectively, without leakage post-selection (data collected on a different day than that of Fig.~\ref{fig:decoder_defect_improvements}). These data compare favorably to other experiments on heavy-hex devices. Ref.~\onlinecite{sundaresan2022} found heavy-hex code logical errors per round of $\lesssim 6\%$ for $Z$ memory experiments and $\lesssim 12\%$ for $X$. Results from our previous work~\cite{lee2026scalable,harper2025characterising} sat around $3 - 4\%$ logical error per round on carefully chosen regions of the device, running codes of size $3\times 3$ or $5\times 5$. We compare this to our current experiments run with the full monolithic device while achieving similar performance.

\begin{figure}
    \centering
\begin{tikzpicture}
\node at (0,0) {\includegraphics[width=0.99\linewidth]{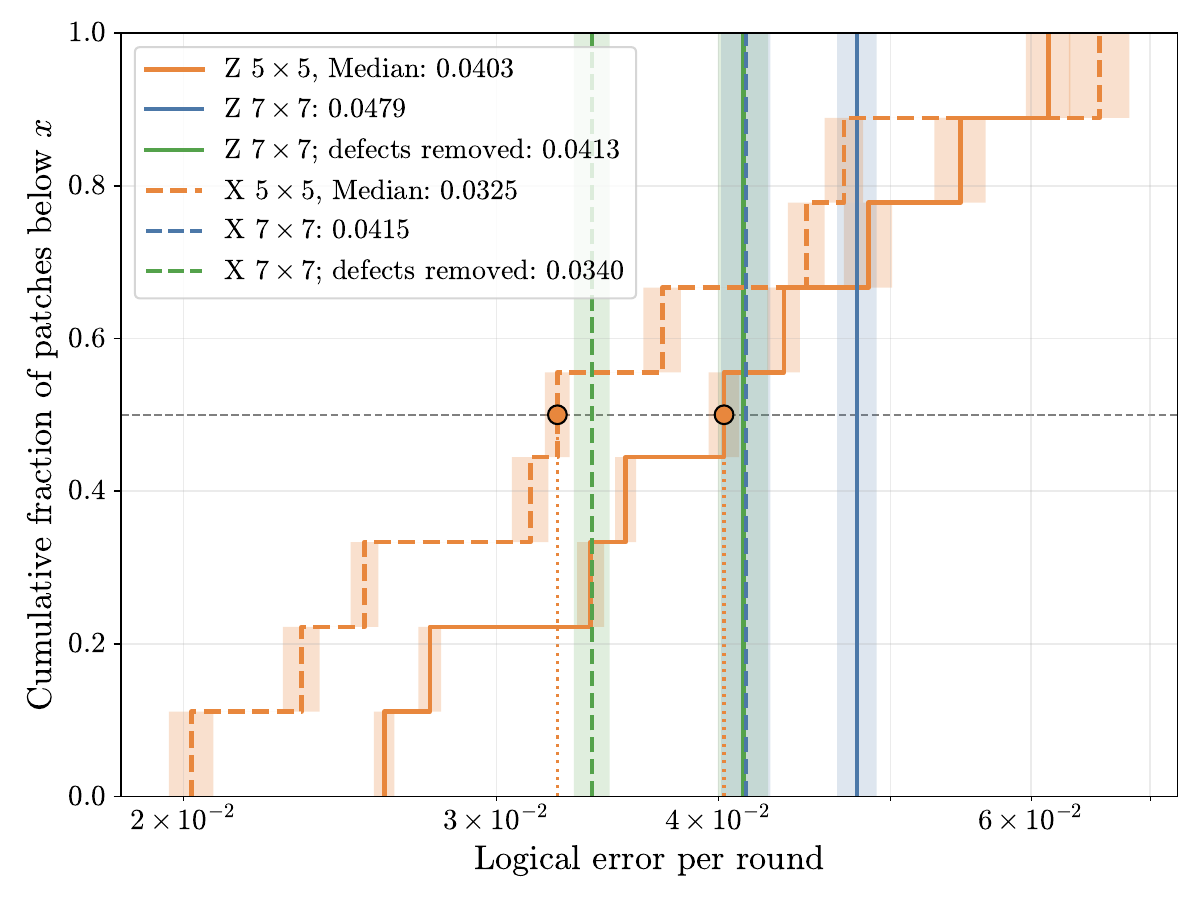}};
\node at (0,-6.5) {\includegraphics[width=\linewidth]{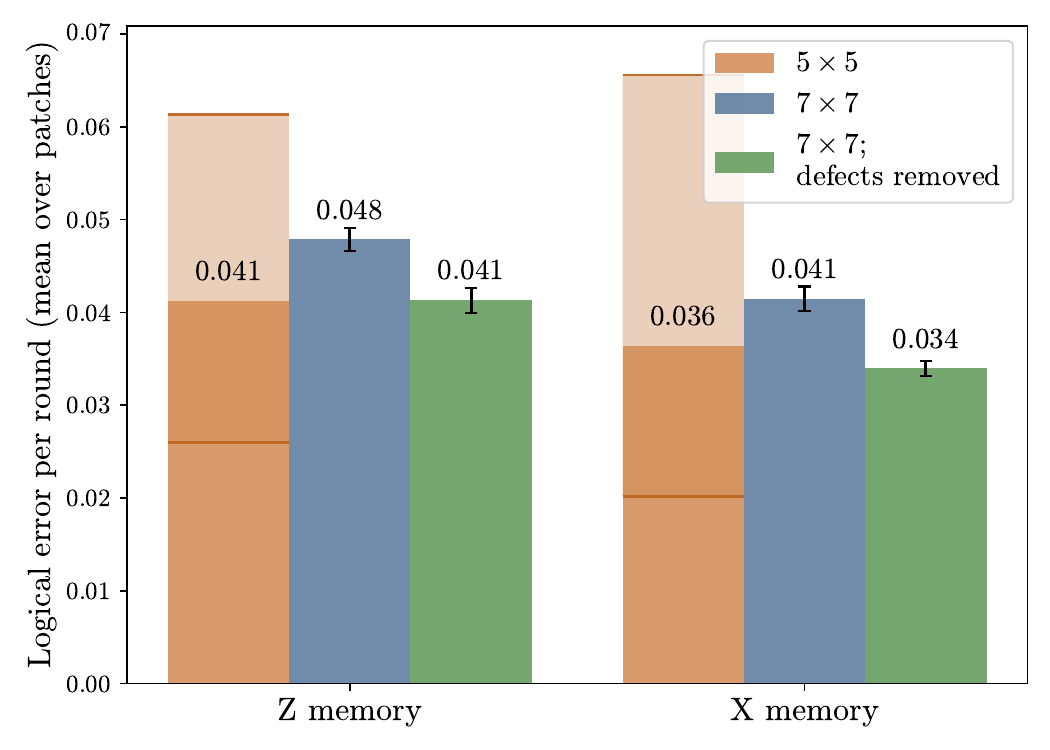}};
\node at (-4,3.5) {(a)};
\node at (-4,-3.2) {(b)};
\end{tikzpicture}
    \caption{\textbf{Scaling memory experiments.} We show the logical error per round for $5\times 5$ and $7\times 7$ codes, the latter both with and without defect mitigation. We use a BeliefMatching decoder and ACES noise information for all decoding. (a) Cumulative distributions of logical error per round, in $Z$ and $X$ memory, for nine $5\times 5$-sized codes covering the footprint of the $7\times 7$ patch. Median logical errors are highlighted with orange circles. We mark the logical error of $7\times 7$ patches both with (green) and without (blue) defect mitigation, in both bases. Shaded regions indicate the uncertainty of that patch's logical error per round. (b) Comparison of mean logical error over patches of size $5\times 5$ and $7\times 7$, both with and without defect mitigation, in both bases. Shaded regions around the bars for $5\times 5$ columns indicate standard errors of the mean over the distribution of patches. Error bars for $7\times 7$ columns indicate uncertainties on logical errors. Data was collected on 2026/06/19.}
    \label{fig:d5_d7_patches}
\end{figure}

\begin{figure*}
    \centering
    \begin{tikzpicture}
    \node at (0,0) {\includegraphics[width=0.53\textwidth]{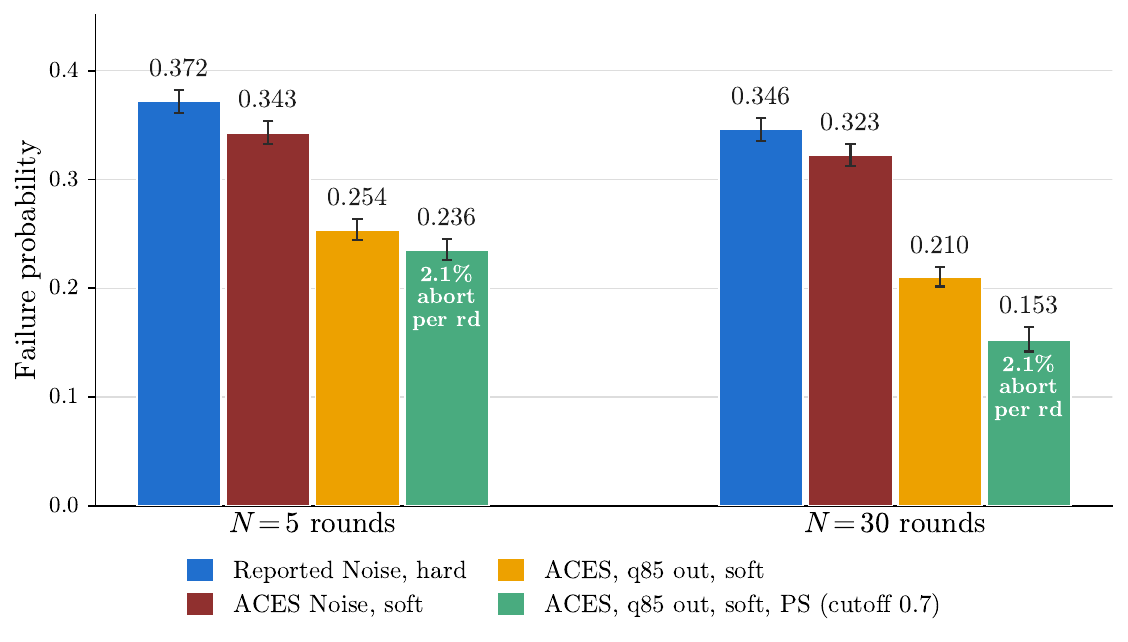}};
    \node at (9,0) {\includegraphics[width=0.44\textwidth]{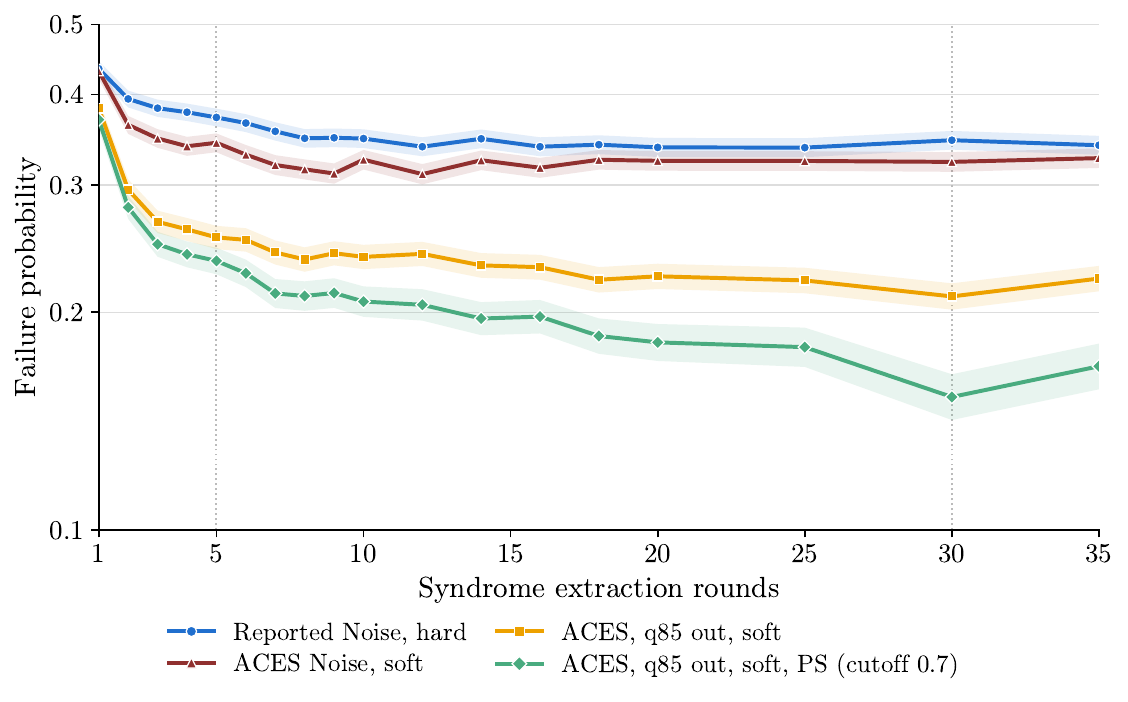}};
    \node at (-4.5,2.5) {(a)};
    \node at (5,2.5) {(b)};
    \end{tikzpicture}
    \caption{\textbf{Stability experiment results.} Stability experiments are performed as in Fig.~\ref{fig:stability_patch}. (a) We report the total failure probability from stability experiments after $N=5$ and $N=30$ rounds of syndrome extraction. We find successive improvements in stability performance from incorporating ACES noise information (with soft information), defect exclusion, and leakage post-selection (see Table~\ref{tab:noise_models}). We compare to a baseline of a reported noise model with no soft information (blue). We then supply the decoder with soft information and ACES noise information (red). We compare this strategy to one of defect exclusion, excluding the components around qubit 85, and using ACES noise and soft information (orange). Finally, we post-select on shots that are unlikely to include leaked qubits, with a cutoff probability of $0.7$ (light green). For leakage post-selection, we fit the abort probability as a function of rounds to an exponential decay, $p_\text{abort} = 1- e^{-N/\tau_a}$, and report the per-round abort probability, $1-e^{-1/\tau_a}$. (b) We report the stability failure probability for values of $N$, the number of syndrome extraction rounds, between $1$ and $35$. The decoder, code and post-selection conditions used are the same as in (a). Data was collected on 2026/06/01.\label{fig:stability_results}}
\end{figure*}

\paragraph*{Comparison of defect-exclusion strategies ---}In Fig.~\ref{fig:dropout_vs_decoder}, we report the percentage improvement in logical error obtained by excluding different defect combinations (compared to a baseline of no defects excluded; note that data was collected on a different day to that of Fig.~\ref{fig:decoder_defect_improvements}). We compare excluding only the defective components around qubit 46, only those around qubit 85, and excluding both. We find that excluding both sets of defective components results in the lowest logical error per round. While we observe that q85 exclusion is no different to excluding both defects in the $Z$ basis, we observe that $X$ basis memory experiments are improved by excluding both defects.

We also observe that the benefit of excluding defects is retained even when supplying the decoder with very accurate ACES noise information. Indeed, in Fig.~\ref{fig:dropout_vs_decoder}, we compare the percentage improvement obtained by defect exclusion across three different decoder approaches, and observe little difference between using reported and ACES noise, indicating that the defects set a floor in performance that cannot be passed by improving the decoder.
However, using a worse decoder and noise model does result in the largest gains from defect exclusion, of $\sim 40\%$ and $\sim 36\%$ in $Z$ and $X$ bases respectively. These result from using a minimum-weight perfect matching decoder~\cite{higgott2021pymatching} informed using a simple global noise model that was chosen to result in reasonable performance across all days tested (See Section~\ref{app:reported_noise}). Hence, noise-informed decoding can reduce the improvement seen from defect exclusion up to a point, but beyond this, active defect exclusion results in a roughly constant percentage improvement. With an ACES-informed BeliefMatching decoder, we see up to $\sim 28\%$ and $\sim 27\%$ improvement in $Z$ and $X$ bases, respectively. We provide further data used to generate Fig.~\ref{fig:dropout_vs_decoder} in Appendix~\ref{app:further_mem_data}.

As a control, in Appendix~\ref{app:further_mem_data}, we demonstrate that excluding a non-defective component in the same patch worsens performance, indicating that the gains we observe are indeed due to defect exclusion strategies.

\paragraph*{Average performance at break-even when scaling a dynamic compass code from $5\times 5$ to $7\times 7$ ---}To gather information about the current state of \textit{Heron} class quantum processors, we interrogate how far scaled memory experiments are from break-even on \texttt{ibm\_boston}. We present results for both $Z$ and $X$ bases in \cref{fig:d5_d7_patches}. Using BeliefMatching and ACES information, we find that the $7\times 7$ patch with defect exclusion is very close to the median and mean $5\times 5$ patch performance in both bases. Note that we only consider the nine $5\times 5$ patches that lie within the footprint of the $7\times 7$ patch. Indeed, in the $X$ basis, defect exclusion allows the $7\times 7$ patch to beat the average $5\times 5$ patch logical error per round. However, we observe that the average $3\times 3$ code still outperforms both larger code sizes, owing to many small codes that can avoid all underperforming components (see Appendix~\ref{app:further_mem_data}).

\subsection{Stability Experiments and Fault-Tolerant Logic}\label{sec:stability_results}

\begin{figure}
    \centering
    \includegraphics[width=\linewidth]{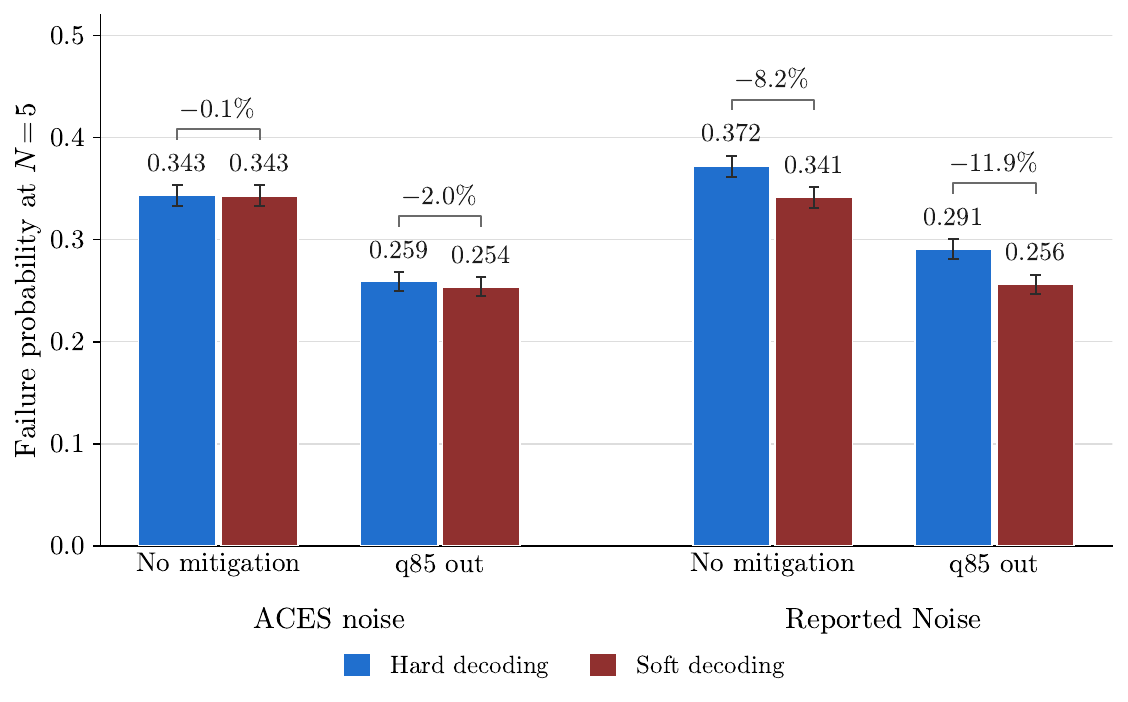}
    \caption{\textbf{Effect of soft information decoding.} We compare the effect of soft information decoding in stability experiments when informing the BeliefMatching decoder with either reported noise or ACES noise information. We show the failure probability at $N=5$ rounds over two defect configurations and two noise models, both with and without soft information. Both when we use no defect mitigation and when we exclude the qubit 85 defect we observe that soft information only makes a statistically significant difference in stability performance when using the reported noise information. When using ACES, no difference is seen when incorporating soft information. The percentage difference in failure probability between hard and soft decoding is reported at the tops of pairs of bars. Data was generated on 2026/06/01.}
    \label{fig:soft_info_stability}
\end{figure}

\begin{figure*}
\begin{tikzpicture}
\node at (0,0) {\includegraphics[width=0.95\linewidth]{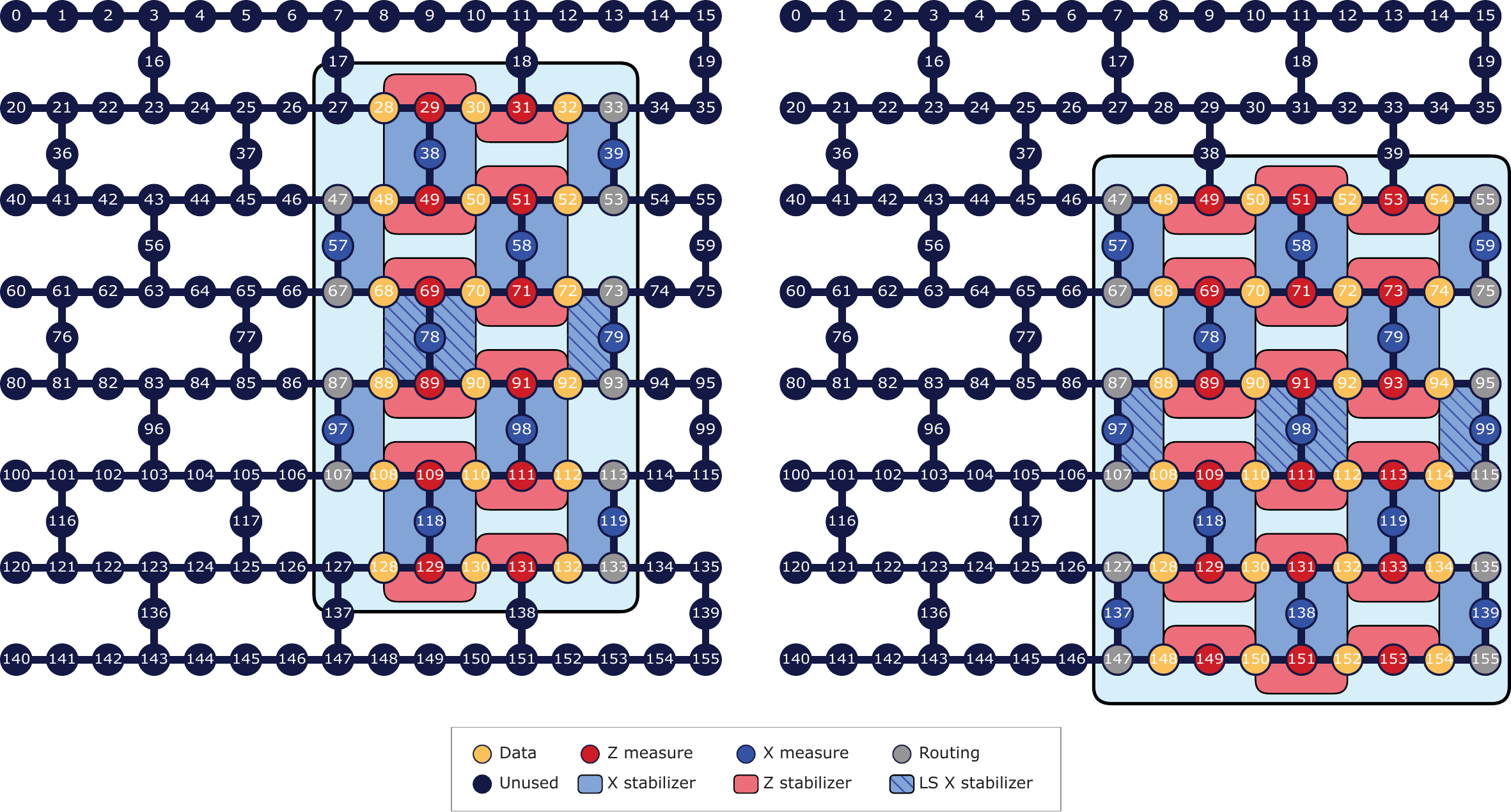}};
\node at (-8.4,5) {(a)};
\node at (0.3,5) {(b)};
\end{tikzpicture}
\caption{\textbf{Lattice surgery on \texttt{ibm\_boston}:} Location of two lattice surgery experiments performed on \texttt{ibm\_boston}. (a) Two $3\times 3$ codes (qubits 28 and 87 in the top-left of the patches) with surgery measurements (blue shaded checks) along the seam gluing the two patches together and thereby reading out the $\bar{X}_1\bar{X}_2$ operator. (b) Two $4\times 3$ codes (qubits 47 and 107 in the top-left of the patches) offer greater protection against $X$ errors during the merge stage.}
\label{fig:lattice_surgery_loc}
\end{figure*}

We quantify the expected performance of fault-tolerant logic gates by performing stability experiments with the dynamic compass code on \texttt{ibm\_boston}~\cite{Gidney2022dual,harper2025characterising,zen2026low}. We construct a $5\times 5$-sized stability patch that covers a large proportion of the device, see Fig.~\ref{fig:stability_patch}(a) for placement. 
We cannot fit a stability patch of the same size as the memory patch, owing to the additional row of $X$-measure qubits required, so we choose a $5\times 5$ stability patch.  

In our stability experiment, the boundary conditions are determined such that the product of all $X$-checks in a given time step is equal to the identity. Hence, in the absence of noise, the product of all $X$-check measurement outcomes at each round of the circuit is $+1$. If, after decoding over $N$ measurement rounds, this observable reads $-1$, we report a logical failure. Measurement-based gates such as lattice surgery require a product of stabilizers to be accurately measured. Stability experiments therefore help to quantify the ability of the device to perform lattice surgery operations, as the observable being tested is a product of stabilizers. The distance of a stability experiment is independent of the size of the patch. Conversely, we expect the circuit distance to scale linearly with the number of syndrome extraction rounds. Hence, when below threshold, we expect to see an exponential decrease in failure probability as we increase the number of rounds. Note that, because mid-circuit resets cannot be performed in \texttt{ibm\_boston}, the time-like distance of our experiments is $\lceil N/2\rceil$, where $N$ is the number of syndrome extraction/measurement rounds~\cite{Geher:2025aa}.

We perform stability experiments on a patch of sufficient size that, for any placement, it will necessarily include some defective components.  As such, our experiments test the effect of defect exclusion on stability experiments. We chose to position the patch as in Fig.~\ref{fig:stability_patch}(a) so that qubit 85 lay close to the middle of the patch. Defects appearing near code boundaries can only be excluded by deforming the boundaries around the defects, thereby resulting in a smaller patch, which itself results in better stability performance~\cite{Gidney2022dual}. The placement of the patch in our experiments therefore allowed us to test the effect of defect exclusion on stability, without changing the size of patch used. For this reason, even though qubit 46 was also covered by the patch, we did not exclude it. In Fig.~\ref{fig:stability_patch}(b), we show the measurement sequence for the defect-excluded stability patch.

In Fig.~\ref{fig:stability_results}, we compare stability performance across four decoding strategies similar to the $7\times 7$ memory experiments. In stability experiments, however, we also employ soft measurement information in some of our strategies; see Table~\ref{tab:noise_models}.
In Fig.~\ref{fig:stability_results}(a), we compare the failure probability after $N=5$ measurement rounds and $N=30$ rounds for each method. 
The overall improvements from a combination of ACES-noise decoding, soft information, and defect exclusion are striking.
Firstly considering no post-selection, at $N=5$, we see an improvement in the failure probability from $0.372$ to $0.254$, giving a reduction in overall failure probability of $32\%.$  At $N=30$, we see an improvement from $0.346$ to $0.21$, giving a reduction of $39\%$. 
When combining leakage post-selection with the above, we see the largest improvements, of up to $56\%$ at $N=30$. However, leakage post-selection on the full patch was not sufficient to reproduce the effects of defect exclusion (see Appendix~\ref{app:more_stability_data}).

Crucially, the combination of defect exclusion and leakage post-selection produces clear error suppression up to $N=30$ rounds. Figure~\ref{fig:stability_results}(b) illustrates failure probability as a function of the number of rounds. Without defect exclusion, the failure probability is suppressed initially by increasing the number of syndrome extraction rounds, but plateaus after around 12 rounds. With defect mitigation excluding defective components around qubit 85, we see not only an improvement of the failure probability after a few rounds but we observe suppression up to $30$ syndrome extraction rounds. We note that Pauli noise simulations do not capture this plateau phenomenon in stability experiments, and we attribute it to leakage in the device.

The stability experiments we perform here are large when compared with other recent experiments, in terms of the number of check measurements per round (16 for the patches in Fig.~\ref{fig:stability_patch}, compared to 12 in Ref.~\cite{drouet2026defects}, 6 in Ref.~\cite{harper2025characterising}, and 4 in Ref.~\cite{caune_demonstrating_2024}). Using a $5\times 5$ patch for our stability experiment, our techniques achieve stability failure probabilities as low as $15.3\%$, which is comparable to previous results ($14.7\%$ in Ref.~\cite{harper2025characterising}, $12.6\%$ with defect-exclusion in Ref.~\cite{drouet2026defects}, $17.6\%$ in Ref.~\cite{caune_demonstrating_2024}). Indeed, all of these previous examples of stability experiments were conducted on smaller patches where we expect performance to be better.

\paragraph*{Soft information ---}We also consider the effect of soft information independently on decoder accuracy. From IQ measurement data, we obtain probabilities that each qubit was in state $0$, $1$, or leaked. When decoding with soft information, we set the prior probability of a measurement error in the decoder to be $\min(p_0, p_1)/(p_0 + p_1)$. Figure~\ref{fig:soft_info_stability} compares the effect of soft vs.\ hard decoding using ACES and reported noise models. 
When using ACES noise, we observe negligible improvement from soft information, but when using reported noise, soft information produces a statistically significant improvement, making the results indistinguishable from the ACES noise results. 
This indicates that for stability experiments, soft information from IQ-plane data may be sufficient for achieving high accuracy decoding, even when a detailed noise model is unavailable.

\subsection{Lattice Surgery}
\label{sec:lattice_surgery}
To evaluate the performance of the device across memory and logic simultaneously, we perform lattice-surgery experiments. We use these to prepare logical Bell states and perform logical state teleportation. Through combinations of pre-selection and post-selection filters, we achieve logical Bell fidelities and teleportation success probabilities over $97\%$. We report logical Bell state preparation experiments here, and teleportation experiments in Appendix~\ref{app:teleport}.

Figure~\ref{fig:lattice_surgery_loc} shows the placement of two surgery experiments: we consider two $3\times 3$ codes (Fig.~\ref{fig:lattice_surgery_loc}(a)) and two $4\times 3$ codes (Fig.~\ref{fig:lattice_surgery_loc}(b)). This code positioning enables lattice surgery but avoids any significantly under-performing couplers or qubits.
For the $3 \times 3$ surgery experiment we use the standard heavy-hex code measurement schedule~\cite{Chamberland2020topological}, where all the Pauli-Z checks are measured simultaneously in every round of measurements. In this case, due to the dimensions of the experiment, the Pauli-$X$ instantaneous stabilizers are weight-$6$. As such, we gain little by using the dynamic compass code schedule to further decrease the weight of the stabilizers. The gains from more frequent $Z$-check measurement outweigh the cost of slightly larger-weight $X$ stabilizers. 
The full four-step sequence of the dynamic compass code is used for the $4\times 3$ lattice surgery. Owing to the greater $X$-distance of the $4\times 3$ codes, this lattice surgery experiment should offer greater protection against logical $Z$ failures in memory during the merge step.

To characterize the surgery experiments, we 
calculate lower and upper bounds of the logical Bell fidelity. 
We do this by measuring the logical Bell-state correlators $\langle\bar{Z}_1\bar{Z}_2\rangle$ and $\langle \bar{X}_1\bar{X}_2\rangle$ from logical qubit readouts after the lattice surgery. Data qubits are initialized in $\ket{0}$ and the lattice surgery is performed for $r=5$ rounds (this maintains a time-like distance of 3 for circuits without resets) to read out the $\bar{X}_1\bar{X}_2$ observable (product of blue shaded plaquettes in Fig.~\ref{fig:lattice_surgery_loc}), obtaining outcome $m_X$. Subsequently, the patches are split and one more round of syndrome extraction is performed for both separated codes. Then, data qubits are measured out either in the $Z$ basis to determine the $\bar{Z}_1\bar{Z}_2$ observable or in the $X$ basis to determine $\bar{X}_1\bar{X}_2$ again. We then perform decoding with BeliefMatching, 
with observables set to either the readout outcomes of $\bar{Z}_1\bar{Z}_2$ or the product of $m_X$ with the readout of $\bar{X}_1\bar{X}_2$.
We are unable to measure $\bar{Y}_1\bar{Y}_2$ fault-tolerantly.
However, we can estimate bounds on the Bell fidelity using $\frac{1}{2}(\langle m_X \bar{X}_1\bar{X}_2\rangle + \langle\bar{Z}_1\bar{Z}_2\rangle)\leq F_{\rm Bell} \leq \frac12(1+\min\{\langle m_X \bar{X}_1\bar{X}_2\rangle, \langle \bar{Z}_1\bar{Z}_2\rangle)$ (See Section~\ref{sec:bell_fidelity}).

\begin{figure*}
\begin{tikzpicture}
        \node at (-5.8,0) {\includegraphics[width=0.61\textwidth]{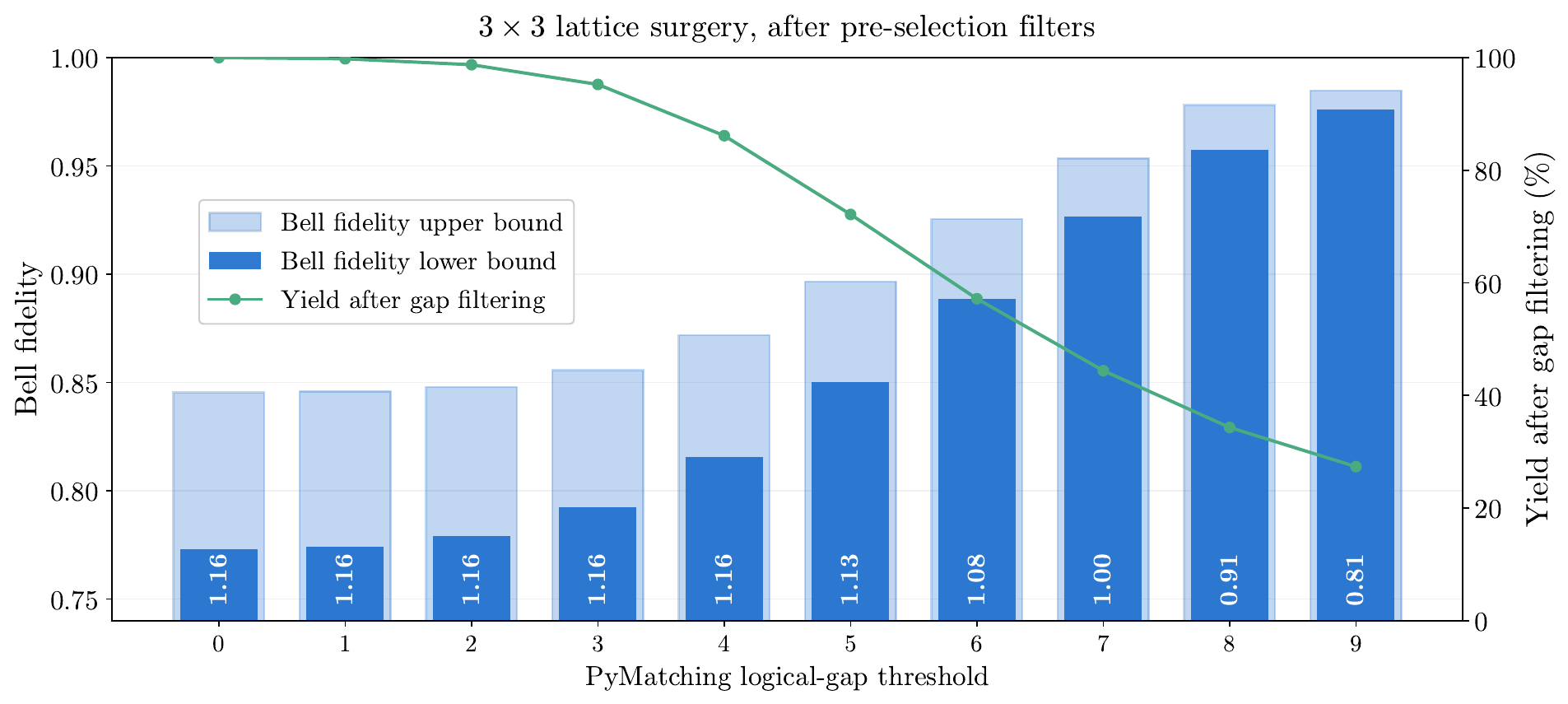}};
        \node at (-10.8,2.5) {(a)};
        \node at (3,0) {\includegraphics[width=0.33\textwidth]{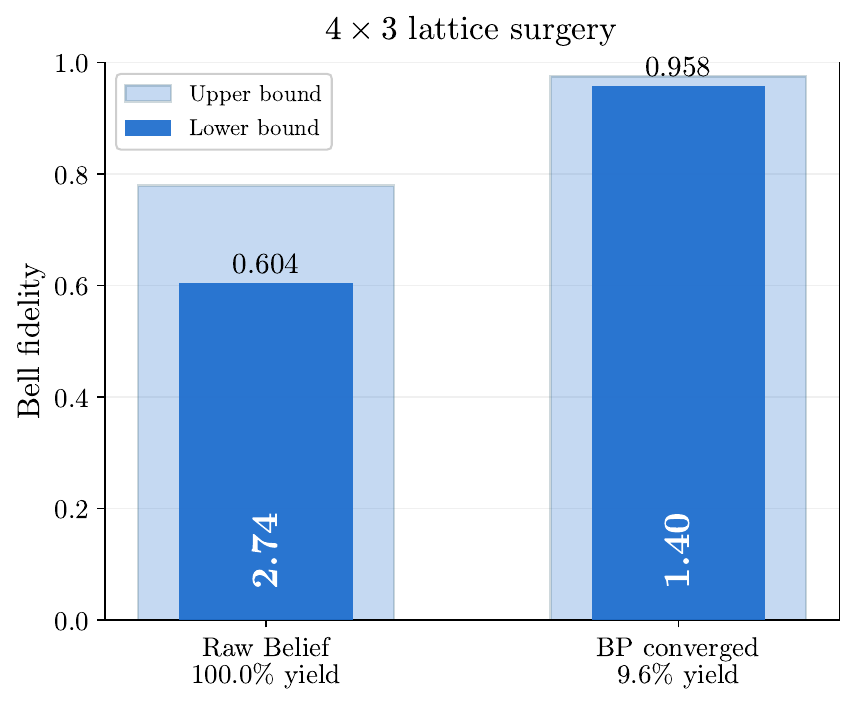}};
        \node at (0.4,2.5) {(b)};
        
\end{tikzpicture}
\caption{\textbf{Logical Bell fidelities from lattice surgery.} (a) Bell fidelity for the $3\times 3$ lattice surgery experiment shown in \cref{fig:lattice_surgery_loc}(a). We initially pre-select on shots that satisfy both: (i) no leakage detected; and (ii) only trivial syndromes for the initial stabilizers observed (see \cref{sec:prefilters}). This indicates that the code is prepared in the code space before the lattice surgery is attempted. Five rounds of lattice surgery are performed, before the logical qubits are separated. We then perform a split of the two patches and finally implement a last round of syndrome extraction before the data qubits are read out in the appropriate basis. Further post-selection is then carried out on a per-shot basis depending on the logical gap calculated by a PyMatching decoder. Final decoding is then performed using the BeliefMatching decoder. The x-axis shows the logical gap thresholds evaluated with an exclusive decoder (shots with a logical gap smaller than this are rejected). Also plotted are the post-selection yields, calculated as a percentage of the shots that have already passed the pre-selection filters. Printed on the bars are the average number of detector violations per round. The lighter shaded (solid) bars show the upper (lower) bound of the estimated Bell Fidelity. (b) Logical Bell fidelity for the $4\times 3$ lattice surgery experiment from \cref{fig:lattice_surgery_loc}(b), with $4\times 3$-sized code patches. Belief propagation convergence is used to perform post-selection on shots: if the belief propagation round converged after 20 iterations, the shot was accepted. Left shows the results without this filter, right the results with the filter. In both cases, no pre-selection filter was applied. All data was collected on 2026/07/17.}\label{fig:lattice-main}
\end{figure*}

\begin{table}
\caption{Lattice surgery Bell fidelity upper and lower bounds with and without pre-selection.}
\label{tab:lattice_surg_pre_sel}
\renewcommand{\arraystretch}{1.5}
\begin{ruledtabular}
\small
\begin{tabular}{lll}
\parbox[t]{0.3\columnwidth}{}& \parbox[t]{0.3\columnwidth}{\raggedright $3\times 3$ surgery} & \parbox[t]{0.3\columnwidth}{\raggedright $4\times 3$ surgery} \\
\hline
\parbox[t]{0.3\columnwidth}{\raggedright No pre-selection} & \parbox[t]{0.3\columnwidth}{\raggedright $58.6\% - 73.5\%$} & \parbox[t]{0.3\columnwidth}{\raggedright $60.4\% - 78.4\%$} \\
\hline 
\parbox[t]{0.3\columnwidth}{\raggedright Full pre-selection} & \parbox[t]{0.3\columnwidth}{\raggedright $77.3\% - 84.5\%$} & \parbox[t]{0.3\columnwidth}{\raggedright $76.6\% - 87.3\%$}
\end{tabular}
\end{ruledtabular}
\end{table}

Without any filtering, the logical Bell fidelity bounds are given in the first row of~\cref{tab:lattice_surg_pre_sel}. 
In teleportation experiments, we initially observe an average success probability across all $X$ and $Z$ basis states of $79.6\%$. To improve these results we implemented slightly different strategies for the different experiments.

\paragraph*{$3\times 3$ surgery ---} We perform pre- and post-selection as follows.  First, we pre-select on shots where we have high confidence that the initial state does not exhibit any leakage, and where we were correctly projected into the initial code state (see~\cref{sec:prefilters}).  In addition, we consider logical gap post-selection using an exclusive decoder~\cite{Smith2024,English_2025,Lee2026PS,chen2026} as follows: we compute a PyMatching confidence score for each shot, given by the difference in minimum matching weight between the decoder’s preferred logical sector and the best correction constrained to the opposite logical sector. We only keep shots whose gap exceeded a chosen threshold. These post-selected shots are then decoded using a BeliefMatching decoder. 

\Cref{fig:lattice-main}(a) shows the results of applying pre-selection filters, together with the effect of using post-selection with an increasing logical-gap threshold, for the $3\times 3$ lattice surgery experiment.
The Bell fidelity bounds obtained using pre-selection are given in the second row of~\cref{tab:lattice_surg_pre_sel}. This strategy filtered out $86.7\%$ of the experiment starting attempts.
Logical gap post-selection then improved these fidelities further. For example, rejecting shots with a logical gap of less than 8 provides  improvement of the logical Bell fidelity lower/upper bound to $95.8\%-97.8\%$ with a relative yield of $34.3\%$ (the total yield after all filters is $4.55\%$). With a threshold of 9, we can achieve Bell fidelities above $97\%$, with a post-selection yield of $27\%$. For teleportation experiments, only using pre-selection filters produced an average success probability of $87.9\%$, while performing pre- and post-selection, we obtain an average success probability of $98.25\%$ with a relative yield of $45.9\%$ (see Appendix~\ref{app:teleport}).

In order to determine whether we are still in an error-correction regime, or if we are simply post-selecting on shots with no detectors firing, we calculate the average number of detector violations per-round after post-selection. At a logical gap threshold of $8$, the average per-round detector violations only drops from $1.16$ to $0.91$, indicating that we are still actively correcting many errors over all shots performed. We report further data, showing the effect of all individual pre-filters and post-selection, in Appendix~\ref{app:teleport}.

We are not aware of results that demonstrate this high a fidelity when performing lattice surgery between two error-correcting ($d>2$) codes in superconducting architectures. For example, Ref.~\onlinecite{besedin_lattice_2026} report post-selected Bell fidelities of up to $78\%$ with $6.3\%$ shot yield when performing lattice surgery between two $d=3$ repetition codes. Meanwhile, Ref.~\onlinecite{wang2026latticesurgery} reports up to $92.5\%$ with $0.5\%$ shot yield for lattice surgery between two $d=3$ surface codes.

\paragraph*{$4\times 3$ surgery ---}When using the combination of pre-selection filters on the $4\times 3$ surgery results, we obtain a yield of $10.34\%$ and a Bell fidelity lower bound slightly worse than the $3\times 3$ results under the same conditions (second row of \cref{tab:lattice_surg_pre_sel}).
When adding post-selection to this experiment, we found little benefit from using the PyMatching confidence score as a post-selection filter. This is because increasing logical gap threshold eventually decreased Bell fidelity estimates; the maximum attainable Bell fidelity lower bound was less than $91.5\%$. 
Instead, we use the BeliefMatching decoder as a confidence filter.
We only kept shots where the belief-propagation stage converged after 20 iterations, computing the logical correction on that accepted subset.
With this post-selection filter we found that neither the leakage pre-filter nor the code-state preparation filter improved the results. 
It appears that conditioning acceptance on the ability of the belief-propagation stage to converge already removes such problematic runs.
Using this filter we were able to achieve a logical Bell fidelity range of $95.8\%-97.5\%$ with a yield of $9.64\%$ and an average per-round detector activity of 1.4 (see \cref{fig:lattice-main}(b)).

\section{Discussion}\label{sec:Discussion}

In this work, we have demonstrated key elements of quantum error correction and fault tolerance with the dynamic compass code on a heavy-hex lattice of 156 superconducting qubits, by running high-fidelity memory, stability and lattice surgery experiments. We combine noise-informed decoding, defect exclusion, and leakage post-selection to achieve significant improvements in memory and stability, and a collection of pre- and post-selection techniques to achieve large improvements in lattice surgery. 
Our results reveal key lessons for future fault-tolerant devices.

We have seen that leaky and under-performing components can severely limit error-correction performance. 
We analyze the role of leakage in our experiments by post-selecting on runs without any detected leakage. This leakage post-selection consistently improved performance of logical elements. A variety of strategies exist and have been tested for reducing leakage in error-correction experiments~\cite{Aliferis_Terhal_LRU_2007,miao_overcoming_2023,mcewen2021,McEwen2023,PRL_leakage2025,Lacroix2025,Werninghaus2021,Battistel2021,xin2025improvederrorcorrectionleakage}, and these will likely be necessary in future generations of devices.
Defects resulted in bottlenecks to improving performance in our experiments and we overcame this issue by introducing defect exclusion techniques for the dynamic compass code. Large numbers of defects will limit the code distances achievable in manufactured chips, indicating that future devices may need to improve component homogeneity in order to see further large performance gains.

We have also seen very sizable benefits from combining pre- and post-selection filters in lattice surgery, which is of significant relevance to future error-correction experiments.  
The pre-selection filters that we use in lattice surgery experiments are qualitatively different than post-selection techniques. Pre-selection allows us to probe the failure mechanism associated with the logic gate itself, removing those mechanisms associated with initializing in a leaked state of the processor or failing to prepare the code space of one or both of the codes. These filters could in principle be applied before any long experiments are begun, as part of a prepare-until-succeed loop~\cite{Gupta2024}. Hence, these could potentially be important components of future experiments for maximizing observed fidelities. 
Separately, lattice surgery often features in magic state factories, which can effectively make use of post-selection~\cite{Litinski2019gameofsurfacecodes}. Performing logical gap post-selection on lattice surgery within such parts of a fault-tolerant architecture could potentially be a scalable~\cite{English_2025,Smith2024,Log_gap_2024,gidney_yoked_2025,meister2024efficientsoftoutputdecoderssurface} method to further boosting resource state fidelities with the additional advantage of aborting low-confidence shots early. 

Finally, we observe that good noise characterization, accurate decoding and measurement soft information, while improving performance up to a degree, are ultimately insufficient to overcome the deleterious effects of defects. However, we have not investigated decoders with higher accuracy than BeliefMatching~\cite{higgott2023improved}. 
Neural network decoders~\cite{Torlai_2017,Bausch2024,Varbonov2025,senior2026scalablerealtimeneuraldecoder} or most-likely error decoders~\cite{beni2025tesseractsearchbaseddecoderquantum} could be investigated in future work to see if these could potentially overcome the need to remove certain under-performing components.
We note that the gains seen from using soft information in stability experiments disappear when using ACES-noise information. This could be practically useful for future experiments, since collecting ACES data can be costly in terms of device time and classical processing time required. However, it would be important to know in what regimes these findings hold.

Future generations of quantum processors will need to improve component homogeneity, to avoid the need for excessive defect exclusion, and will require fast, multi-level resets or other techniques to actively remove leakage. They will need to maintain high component fidelities and homogeneity while scaling to larger sizes of chip. The use of defect exclusion strategies that are far less costly in terms of distance and fidelity could be achieved by considering designs with more connectivity between qubits~\cite{drouet2026defects}. There will in general be an interplay between chip homogeneity, component fidelity, and fabrication yield, with greater qubit connectivity increasing tolerance to defects. We envisage that these future devices will allow for further key error-correction primitives to be demonstrated, such as magic state injection and distillation, a greater array of Clifford operations, and real-time accurate decoding.  Finally, future quantum processors with a large number of logical qubits will require even more qubits and/or other innovations such as long-range connectivity that will allow for larger-distance quantum codes and lower logical error rates.

\section{Methods}\label{sec:Methods}

\subsection{Noise models, defect exclusion, and leakage post-selection}
\label{app:reported_noise}

\begin{table*}
\caption{\label{tab:noise_models}\textbf{Noise-informed strategies.} We compare the noise model, syndrome extraction circuit, and degree of post-selection used in each experiment.}
\begin{ruledtabular}
\small
\begin{tabular}{lll}
Noise model
 & \parbox[b]{0.50\textwidth}{\centering Memory}
 & \parbox[b]{0.22\textwidth}{\centering Stability} \\
\hline
\noalign{\vskip \rowgap}
\parbox[t]{0.16\textwidth}{\raggedright Reported Noise\\ (blue data)}
 & \parbox[t]{0.50\textwidth}{\raggedright Set the prior probabilities of each noise process using IBM-reported characterization data for \texttt{ibm\_boston}.
The construction of this Pauli noise model is described below.}
 & \parbox[t]{0.22\textwidth}{\raggedright Same as memory.} \\
\noalign{\vskip \rowgap}
\parbox[t]{0.16\textwidth}{\raggedright ACES Noise\\ (red data)}
 & \parbox[t]{0.50\textwidth}{\raggedright A Pauli noise model extracted using ACES~\cite{ACESgithub,Hockings_QuantumACES_jl_design_noise_2025}, which has layer-specific reconstructed noise information. }
 & \parbox[t]{0.22\textwidth}{\raggedright As with memory, but also using soft measurement information.} \\
\noalign{\vskip \rowgap}
\parbox[t]{0.16\textwidth}{\raggedright ACES, q46 and q85 excluded\\ (orange data)}
 & \parbox[t]{0.50\textwidth}{\raggedright Same as ACES Noise, using the syndrome extraction circuit that excludes qubits $46$ and $85$, see Fig.~\ref{fig:Main_code_patch}(b).}
 & \parbox[t]{0.22\textwidth}{\raggedright As with memory, but only excluding qubit $85$ and using soft information.} \\
\noalign{\vskip \rowgap}
\parbox[t]{0.16\textwidth}{\raggedright ACES, q46 and q85 excluded, PS at cutoff $p$\\ (light green data)}
 & \parbox[t]{0.50\textwidth}{\raggedright Same as ACES, excluding q46 and q85.  Additionally, shots that report greater than $p=0.9$ probability that any measured qubit has leaked, using IQ data, are discarded.}
 & \parbox[t]{0.22\textwidth}{\raggedright As with memory, but only excluding qubit $85$, using soft information, and with cutoff probability $p=0.7$.} \\
\noalign{\vskip \rowgap}
\end{tabular}
\end{ruledtabular}
\end{table*}

In Table~\ref{tab:noise_models}, we provide a summary of the columns in Figs.~\ref{fig:decoder_defect_improvements} and \ref{fig:stability_results}. We use either a reported noise model (explained below) or an ACES noise model (explained in Refs.~\onlinecite{lee2026scalable,hockings2024,hockings2025improvingerrorsuppressionnoiseaware}), and we exclude defects (explained in Section~\ref{sec:defect_exclusion_details}), and perform leakage post-selection (explained in detail in Ref.~\onlinecite{lee2026scalable}).

For some decoding, we report using a ``simple" global noise model. This was chosen to perform comparably to other noise models regardless of the day of data collection. In Table~\ref{tab:simple_noise_params}, we provide the parameters and locations in the circuit in which noise was introduced in that noise model.

\begin{table}
\caption{Simple noise model parameters.}
\label{tab:simple_noise_params}
\begin{ruledtabular}
\small
\begin{tabular}{ll}
\parbox[t]{0.5\columnwidth}{\raggedright Noise location} & \parbox[t]{0.45\columnwidth}{\raggedright Noise channel and strength} \\
\hline
\parbox[t]{0.5\columnwidth}{\raggedright Single-qubit gates} & \parbox[t]{0.45\columnwidth}{\raggedright $\texttt{DEPOLARIZE1}(0.0005)$} \\
\parbox[t]{0.5\columnwidth}{\raggedright Two-qubit gates} & \parbox[t]{0.45\columnwidth}{\raggedright $\texttt{DEPOLARIZE2}(0.003)$} \\
\parbox[t]{0.5\columnwidth}{\raggedright Idling during two-qubit gates} & \parbox[t]{0.45\columnwidth}{\raggedright $\texttt{DEPOLARIZE1}(0.0005)$} \\
\parbox[t]{0.5\columnwidth}{\raggedright Idling during measurements} & \parbox[t]{0.45\columnwidth}{\raggedright $\texttt{DEPOLARIZE1}(0.01)$} \\
\parbox[t]{0.5\columnwidth}{\raggedright Pre- and post-measurement error} & \parbox[t]{0.45\columnwidth}{\raggedright $\texttt{X-ERROR}(0.01)$} \\
\parbox[t]{0.5\columnwidth}{\raggedright Classical measurement error} & \parbox[t]{0.45\columnwidth}{\raggedright $\texttt{M}(0.01)$} \\
\end{tabular}
\end{ruledtabular}
\end{table}

We now explain how the reported noise models were constructed. For the memory and stability experiments, 
we construct the reported-noise baseline from the IBM calibration snapshot supplied with the experimental dataset, obtained on the same date as data collection.
Following the calibration-snapshot approach of Ref.~\cite{lee2026scalable}, we use the reported error rates to parameterize a circuit-level Pauli noise model that supplies prior probabilities to the decoder.
We retain the values for individual physical qubits and couplers, including those associated with defective components, rather than averaging them over the device.
Each retained parameter is fixed across circuit layers and experimental shots.
This model therefore captures spatial variations in the reported noise, while omitting the layer-specific characterization provided by ACES.

\paragraph*{Gate and idle errors.}
The fields used from the snapshot are listed in Table~\ref{tab:reported_noise_fields}.
For a single-qubit gate acting on qubit $q$, we use its \texttt{RX error} as the probability $p$ in the depolarizing channel
\begin{equation}
\mathcal{D}_{1,p}(\rho)=(1-p)\rho+\frac{p}{3}\sum_{P\in\{X,Y,Z\}}P\rho P.
\label{eq:reported_depolarizing_one}
\end{equation}
For a two-qubit gate on a coupler connecting qubits $q$ and $q'$, we use the corresponding \texttt{CZ error} in
\begin{equation}
\mathcal{D}_{2,p}(\rho)=(1-p)\rho+\frac{p}{15}\sum_{P\in\mathcal{P}_2\setminus\{I\otimes I\}}P\rho P,
\label{eq:reported_depolarizing_two}
\end{equation}
where $\mathcal{P}_2=\{I,X,Y,Z\}^{\otimes 2}$.
Coupler error rates are indexed by the unordered pair of physical qubits.
These channels are attached to the corresponding ideal operations in the syndrome-extraction circuit.
An idling qubit receives the channel in Eq.~\eqref{eq:reported_depolarizing_one}, with $p$ given by its \texttt{ID error}, at each idle location specified by the circuit schedule, including when it remains idle while other qubits undergo mid-circuit measurements.
After the preprocessing described below, the reported values are used directly as the Pauli error probabilities $p$ in these equations.

\begin{table}[t]
\caption{Reported fields used to construct the baseline noise model.
The gate and idle fields set depolarizing probabilities; the measurement fields set classical readout outcome flip probabilities.}
\label{tab:reported_noise_fields}
\begin{ruledtabular}
\small
\begin{tabular}{ll}
\parbox[t]{0.30\columnwidth}{\raggedright Snapshot field} & \parbox[t]{0.61\columnwidth}{\raggedright Circuit assignment} \\
\hline
\parbox[t]{0.30\columnwidth}{\raggedright \texttt{RX error}} & \parbox[t]{0.61\columnwidth}{\raggedright Single-qubit gate on the reported qubit} \\
\parbox[t]{0.30\columnwidth}{\raggedright \texttt{CZ error}} & \parbox[t]{0.61\columnwidth}{\raggedright Two-qubit gate on the reported coupler} \\
\parbox[t]{0.30\columnwidth}{\raggedright \texttt{ID error}} & \parbox[t]{0.61\columnwidth}{\raggedright Idle location on the reported qubit} \\
\parbox[t]{0.30\columnwidth}{\raggedright \texttt{MEASURE error}} & \parbox[t]{0.61\columnwidth}{\raggedright Longer pulse, used for final data-qubit readout} \\
\parbox[t]{0.30\columnwidth}{\raggedright \texttt{MEASURE\_2 error}} & \parbox[t]{0.61\columnwidth}{\raggedright Shorter pulse, used for mid-circuit syndrome readout} \\
\end{tabular}
\end{ruledtabular}
\end{table}

\paragraph*{Measurement errors.}
IBM supports two measurement pulses with different durations, whose reported error rates are listed as \texttt{MEASURE error} for the longer pulse and \texttt{MEASURE\_2 error} for the shorter pulse.
In our circuits, we use the longer pulse for final data-qubit readout and the shorter pulse for mid-circuit syndrome readout.
Each reported measurement error probability $p$ is assigned to an independent classical flip of the recorded outcome, leaving the post-measurement quantum state unchanged.
Note that this assignment differs from the pre-measurement-noise convention used for the calibration baselines in Ref.~\cite{lee2026scalable}.
The reported measurement error alone does not resolve the contributions of quantum errors before measurement and classical assignment errors, so their allocation is a modeling assumption.
When soft information is used, the probabilities of classical measurement errors are updated on a shot-by-shot basis using IQ data; this does not change the gate or idle parameters defined here.

\paragraph*{Unavailable and unit-valued entries.}
We treat zero, missing, and non-finite error entries as unavailable and replace them with the mean of the finite nonzero entries of the same operation type in the snapshot.
For the two-qubit errors, this procedure uses the reported coupler-error entries.
Probabilities are restricted to the interval $[0,1]$.
The data provided by IBM contain entries with an error probability of $1.0$.
We regard these entries as indicating a problem with the characterization process or the reported data, and replace them with the largest finite probability strictly below unity for the same operation type.
This retains a high error prior for the affected component while avoiding an error probability of one.
The replacement is carried out independently for the \texttt{RX}, \texttt{ID}, \texttt{CZ}, \texttt{MEASURE}, and \texttt{MEASURE\_2} fields.

\paragraph*{Scope of the approximation.}
The baseline does not infer additional noise channels from the reported $T_1$ or $T_2$ times, nor does it rescale idle errors using gate or readout durations.
In particular, qubits that remain idle while other qubits are measured use the same reported \texttt{ID error}, even though measurements can take substantially longer than gates.
No separate reset error or post-measurement state error is added, and leakage dynamics and temporal drift are not modeled explicitly.

\subsection{Defect exclusion strategies}\label{sec:defect_exclusion_details}

\begin{figure*}
\begin{tikzpicture}
    \node[anchor=south west, inner sep=0] (img) at (0,0) {\includegraphics[width=0.7\textwidth]{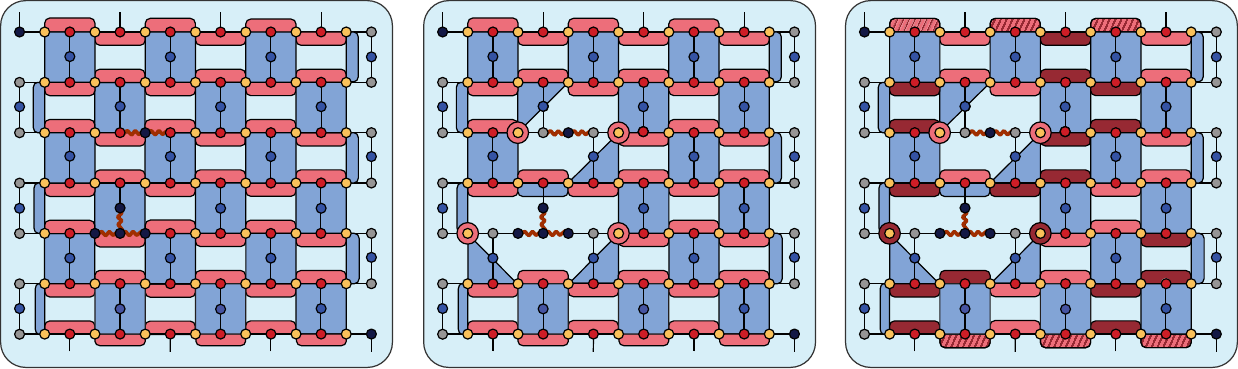}};
    \node[anchor=north west] at (1.68, -0.15) {(a)};
    \node[anchor=north west] at (5.97, -0.15) {(b)};
    \node[anchor=north west] at (10.28, -0.15) {(c)};
\end{tikzpicture}
\caption{Defect exclusion for qubits 46 and 85 in \texttt{ibm\_boston}. Data qubits are shown in yellow, $X$ ($Z$) measure qubits in blue (red), and unused qubits in black. (a) Defective couplers (red wavy lines) to be excluded from the patch and their adjacent qubits (black), which themselves may be defective, are identified. (b) Check operators with support on excluded components are truncated away from those components and now have reduced weight. (c) A check measurement schedule is assigned to minimize the weight of superstabilizers around the defects. Those checks measured in Step 2 (Step 4) of Fig.~\ref{fig:Main_code_patch} are colored dark red (light red), and those checks on the top/bottom boundaries measured in both Steps are highlighted with stripes.\label{fig:defect_exclusion_strat}}
\end{figure*}

Defective components in the Boston device tend to aggregate around particular locations. The couplers around qubits 146, 46 and 85 tend to consistently underperform, while other qubits and couplers have varying levels of performance. We choose to limit defect exclusion to two locations within the code patch so as to avoid reducing the distance of the $7\times 7$ patch to less than 5 in both bases. On all days tested, couplers around qubits 85 and 146 had $CZ$ gate errors that were 2-3 orders of magnitude larger than the median $CZ$ gate error across the device, and the next-worst coupler was consistently neighboring qubit 46. Meanwhile, $T_1$ and $T_2$ times were often poorest for qubits 85, 146 and 46. Specifically, on the date during which the majority of our data was taken (2026/06/01), 
qubit 46 had the worst $T_2$ time ($4.23\mu $s) and the second-worst $T_1$ time behind qubit 85 ($9.38\mu$s). On another data collection date (2026/06/19), qubit 46 had the second-worst $T_1$ time behind qubit 85 ($14.45\mu$s) but a better $T_2$ time relative to other qubits. Owing to the underperforming couplers and qubits in those three regions specified, we chose to consistently treat these three locations as defective and test defect exclusion strategies on them.

We treat these defects in different ways. Qubit 146 lies on the bottom row of the chip. We can place a $7\times 7$ code on the chip while avoiding the bottom row entirely and exclude the defective component in this way. Meanwhile, qubits 46 and 85 lie in the middle of any $7\times 7$ code, and must be excluded by deforming the code. Note that excluding any coupler from the code leaves one qubit with only a single coupler touching it, owing to every coupler touching one valence-two qubit. If this semi-isolated qubit is a data qubit, it is prohibitively challenging to measure multiple checks that have support on that qubit, and hence we further exclude that data qubit from the code. In Fig.~\ref{fig:defect_exclusion_strat}, we show how excluding underperforming couplers (red wavy lines) around qubit 85 (located on a 3-valent vertex) leaves all neighboring qubits semi-isolated (adjacent to only one remaining coupler). We exclude these from the code as well. Meanwhile, excluding qubit 46 and its adjacent couplers does not isolate neighboring qubits, which are located at vertices, so these are retained in the code.

We restrict all checks to the qubits that remain in the code after excluding defective components and semi-isolated qubits. For the defect arrangement considered in our experiments, the result of these restrictions is shown in Fig.~\ref{fig:defect_exclusion_strat}(b). Note that the checks now include single-data-qubit measurements in the $Z$ basis and two- or three-qubit $X$-checks. We then construct a measurement sequence for these checks and form superstabilizers which change over the four steps of the QEC sequence, as shown in Fig.~\ref{fig:Main_code_patch}. We measure all $X$-checks in Steps 1 and 3 as normal for the dynamic compass code, and measure a subset of $Z$-checks in Step 2, with the remaining $Z$-checks measured in Step 4. We always choose $Z$-checks that are in the same row as -- and adjacent to -- a superstabilizer to be measured in a different step to the superstabilizer. This helps maintain low-weight $X$ superstabilizers in the vicinity of defects. The measurement pattern chosen is shown in Fig.~\ref{fig:defect_exclusion_strat}(c). This results in maximum weight 10 $X$ superstabilizers and maximum weight 8 $Z$ superstabilizers (see Fig.~\ref{fig:Main_code_patch}). 

Excluding defects from the code patch changes the distance of the code. In Table~\ref{tab:d7-dropout-distance-checks}, we list the $Z$ and $X$ distances of the code patches with one or both defects excluded from the patch.

\begin{table}
\centering
\begin{tabular}{lcc}
\hline
Circuit & $Z$-distance & $X$-distance \\
\hline
Full patch & 7 & 7 \\
q46 out & 6 & 6 \\
q85 out & 5 & 6 \\
Both out & 5 & 5 \\
\hline
\end{tabular}
\caption{Distances for the the $7\times 7$ dynamic compass code patches with defect mitigation.}
\label{tab:d7-dropout-distance-checks}
\end{table}

\begin{figure}
    \centering
\begin{tikzpicture}
\node at (0,0) {\includegraphics[width=0.95\linewidth]{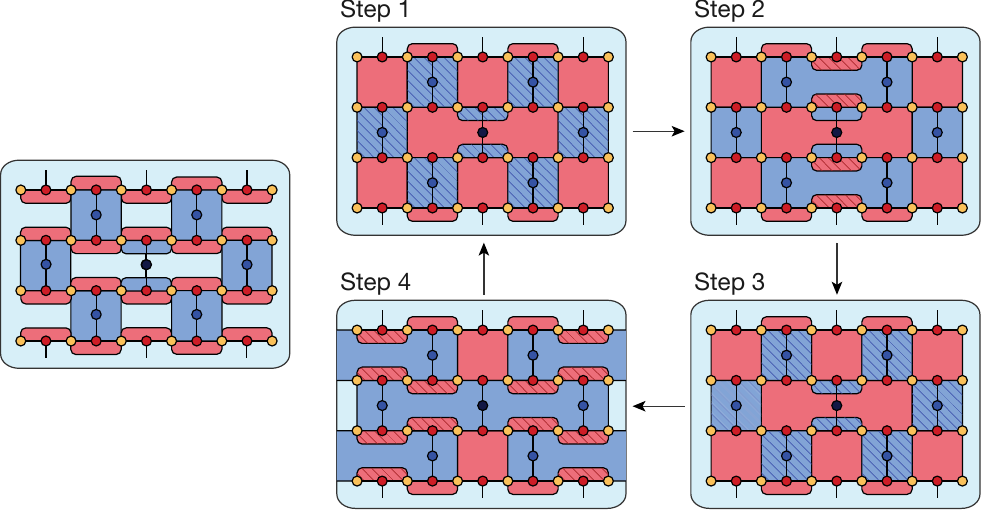}};
\node at (-4,1.2) {(a)};
\node at (-1,2.5) {(b)};
\end{tikzpicture}
    \caption{Exclusion of a defective $X$-measure qubit, shown in black. A small region of dynamic compass code is shown. (a) We show the checks of the new code. The $X$-check that was to be measured at the defect is split into two 2-body checks, shown in blue. (b) The four-step measurement sequence for the dynamic compass code with the defect excluded is shown. Shaded plaquettes indicate measurements made in that step and unshaded regions indicate elements of the instantaneous stabilizer group during that step.}
    \label{fig:vertical_dropout_schedule}
\end{figure}

We have demonstrated how to exclude defective data qubits and $Z$-measure qubits (located on vertices of the lattice). Since we exclude semi-isolated data qubits, we therefore treat defective horizontal couplers as defective data qubits and perform defect exclusion in the same manner. For completeness, we also show the result of excluding a defective $X$-measure qubit (located on vertical edges of the lattice), in Fig.~\ref{fig:vertical_dropout_schedule}. Since the corresponding $X$-check cannot be measured at that defective location, we split it up into two-body $X$-checks, as shown in Fig.~\ref{fig:vertical_dropout_schedule}(a). These newly created $X$-checks result in weight-8 $Z$-superstabilizers appearing in the measurement schedule, as shown in Fig.~\ref{fig:vertical_dropout_schedule}(b). This results in a $Z$-distance of the code which is lowered by 2. Meanwhile, the $X$-distance of the code is maintained. Defective vertical couplers can be treated in a similar way, by splitting up the $X$-check at that location into two-body checks.

The new two-body $X$-checks that appear in the defect-excluded codes (Figs.~\ref{fig:defect_exclusion_strat} and \ref{fig:vertical_dropout_schedule}) can be measured by re-purposing the $Z$-measure qubits that they straddle, shown in red in the figures. However, to maintain a compressed measurement schedule, in which $Z$ and $X$ checks are measured in the same time step (important for achieving high logical fidelities~\cite{harper2025characterising}), we must adopt a different approach, which we outline in the next section.

\subsection{Syndrome measurement circuits}\label{sec:syndrome_mmt}

Here, we provide more details on the syndrome measurement circuits performed during memory and stability experiments. We first describe the circuits used for the regular and defect-excluded dynamic compass code, then comment on detectors for circuits without resets, and finally discuss details of the circuits required for a compressed measurement schedule. 

\subsubsection{Dynamic Compass Code Circuits}

Regular $X$-checks are measured using a depth-8 circuit of CNOT gates that connect the measure qubit (blue in Fig.~\ref{fig:Main_code_patch}) to the four data qubits (yellow) via the $Z$-measure qubits (red) to the North and South of the $X$-measure qubit. This circuit can be understood as four ``via gates", each performing one CNOT between an $X$-measure qubit and a data qubit. The via gates themselves are depth-4 CNOT circuits, although we use circuit identities to lower the overall depth of the $X$-check measure circuit~\cite{harper2025characterising}. A depth-2 CNOT gate circuit is then used to extract the syndrome of a subset of $Z$-checks. Measurements of \textit{all} $Z$ and $X$ measure qubits are then made simultaneously. The CNOT circuits for extracting $X$ and $Z$ check information are described in more detail in Ref.~\onlinecite{harper2025characterising} and can be seen in Fig.~\ref{fig:circuit_patch_overlay}.

Around defects, the checks measured include three-body and two-body $X$-checks, and single-body $Z$-checks. These lower-weight $X$-checks can be measured with subsets of the gates from the four-body check measurement circuit. For two-body checks, we can also simply use a circuit that measures out the parity information on the $Z$-measure qubit in between the two data qubits. Occasionally, we need to employ a different ``Bell pair measurement" method, outlined below. 

We begin all memory experiments with a full set of $Z$-check measurements followed by a full set of $X$-check measurements, which initialises the code space. In all data plots, we count the number of rounds as the number of measurement layers, equivalent to double the number of full four-step syndrome extraction cycles. We start from rounds performed after the setup of the code space.

In Fig.~\ref{fig:circuit_patch_overlay}, we show the circuit used to measure the specific checks shown in Fig.~\ref{fig:defect_exclusion_strat} and thereby implement the memory experiment of Fig.~\ref{fig:Main_code_patch}. Other circuits run in this paper are highly analogous. We show the circuit over two rounds (four steps) of syndrome extraction.

\subsubsection{Detectors}

We form detectors that compare the measurement of certain subsets of $X$ or $Z$ checks. Detectors are defined to be the parity of sets of measurement outcomes, which are always equal to $+1$ in the absence of noise. Since \texttt{ibm\_boston} does not have functionality for mid-circuit resets, we perform memory and stability circuits without resets~\cite{Geher:2025aa,harper2025characterising}, requiring us to modify the definition of detectors to accommodate this. If $M_t$ denotes a deterministic combination of measurement outcomes in round $t$, corresponding to the measurement of an (instantaneous) stabilizer, then in the absence of noise and without resets, $M_{j>t} = M_t$. For measurements of measure qubit $i$, $m^{(i)}_t$, the true outcome of the measurement is $0$ if $m^{(i)}_t = m^{(i)}_{t-1}$, and $1$ if $m^{(i)}_t \neq m^{(i)}_{t-1}$. Hence, we replace each measurement $m_t$ in the detector definitions with the combination $m_t\oplus m_{t-1}$, since this results in detectors that are strictly $+1$ (rather than simply $+1$ or $-1$ deterministically) in the absence of noise. Any measurement made that does not reveal stabilizer information nevertheless can be used to reveal flag information~\cite{Chamberland2020topological}. We compare the outcomes of these flag qubit measurements to their previously measured values and form flag detectors in this way. These appear on boundaries as well as in the bulk of the code, since $Z$-measure qubits are not used to read out check information in half of the measurement rounds.

Using our prescription of adding in the previous measurements (at the same qubits) to detectors can result in unnecessarily high-weight detectors, where we define the weight of a detector as the number of measurement outcomes that feature therein. In particular, when $Z$ measure qubits are not being used for syndrome extraction (e.g., in Steps 2 or 4 of Fig.~\ref{fig:Main_code_patch}), they are nevertheless measured and we use these measurements to form flag detectors. We can use these flag detectors to reduce the weights of other detectors. In particular, the prescription of including previous measurement outcomes in detectors would normally result in weight-8 $Z$-detectors (each measurement from the initial weight-4 detector in the circuit with resets, $m_t^{(1)}m_t^{(2)}m_{t-2}^{(1)}m_{t-2}^{(2)}$, picks up one previous measurement outcome). However two of these measurement outcomes (those at $t-1$ and $t-2$) form a flag detector, since the $Z$-measure qubit is not being used for check readout at time $t-1$. Hence, $m^{(i)}_{t-1}m^{(i)}_{t-2}$ can be removed from the $Z$-detector. In the bulk of the circuit, therefore, $Z$ detectors are maintained as only weight-4, and they compare two $Z$-check outcomes in round $t$ with the same checks' outcomes in round $t-3$. $X$ detectors are formed analogously to detectors in surface codes without resets~\cite{Geher:2025aa,drouet2026defects}: we compare measurements in round $t$ to those in round $t-2$, and $X$ detectors in the bulk also have weight 4.

\subsubsection{Bell pair measurements near defects}\label{sec:bell_pair_mmt}

Around defects, we sometimes need to measure $XX$ and $ZZ$ operators on the same pair of qubits simultaneously. See, for example, the two-body $Z$ and $X$ checks supported on data qubits immediately above and below the defect in Fig.~\ref{fig:vertical_dropout_schedule}(a) (see also the two-body $X$ check in Fig.~\ref{fig:Main_code_patch}(b)). In Steps 3 and 4 of the measurement sequence in Fig.~\ref{fig:vertical_dropout_schedule}(b), the $ZZ$ operators are not measured, and hence the two-body $XX$ operator is measured out using the $Z$-measure qubit, as is usual for two-body operators in the code. We can achieve this without increasing the gate depth of the syndrome extraction circuit. However, in Steps 1 and 2, we need to measure both $ZZ$ and $XX$ operators simultaneously. This is sometimes possible to achieve using measure qubits, but for general defect patterns, it is not always the case that there are sufficient measure qubits available in the vicinity of the defect to perform both two-body measurements simultaneously. Hence, in our circuits, we perform these as Bell-type measurements, measuring out the \textit{data} qubits to learn the check information. 

\begin{align}
&\begin{quantikz}[ampersand replacement=\&,
                  row sep={0.55cm,between origins}, column sep=0.35cm]
  \lstick{$d_1$} \& \meter{$ZZ$}\vqw{2} \& \meter{$XX$}\vqw{2} \& \\
  \lstick{$\ket{0}$} \&     \&     \&  \gate{H}\\
  \lstick{$d_2$} \&     \meter{}            \&      \meter{}           \&
\end{quantikz}
\;\;= \nonumber \\[1.5em]
&\quad \begin{quantikz}[ampersand replacement=\&, row sep={0.55cm,between origins}, column sep=0.3cm]
  \lstick{$d_1$} \& \ctrl{1} \&          \& \ctrl{1} \& \gate{H} \& \meter{} \& \gate{H} \&   \&    \ctrl{1}      \&    \& \\
  \lstick{$\ket{0}$} \& \targ{}  \& \ctrl{1} \& \targ{}  \&          \& \meter{} \& \gate{H} \& \ctrl{1}  \& \targ{} \& \ctrl{1}  \& \\
  \lstick{$d_2$} \&          \& \targ{}  \&          \&          \& \meter{} \&  \&   \targ{}      \&   \&    \targ{}      \&
\end{quantikz}
\label{eqn:Bell_mmt_gadget}
\end{align}

\begin{figure*}
    \centering
    \includegraphics[width=\linewidth]{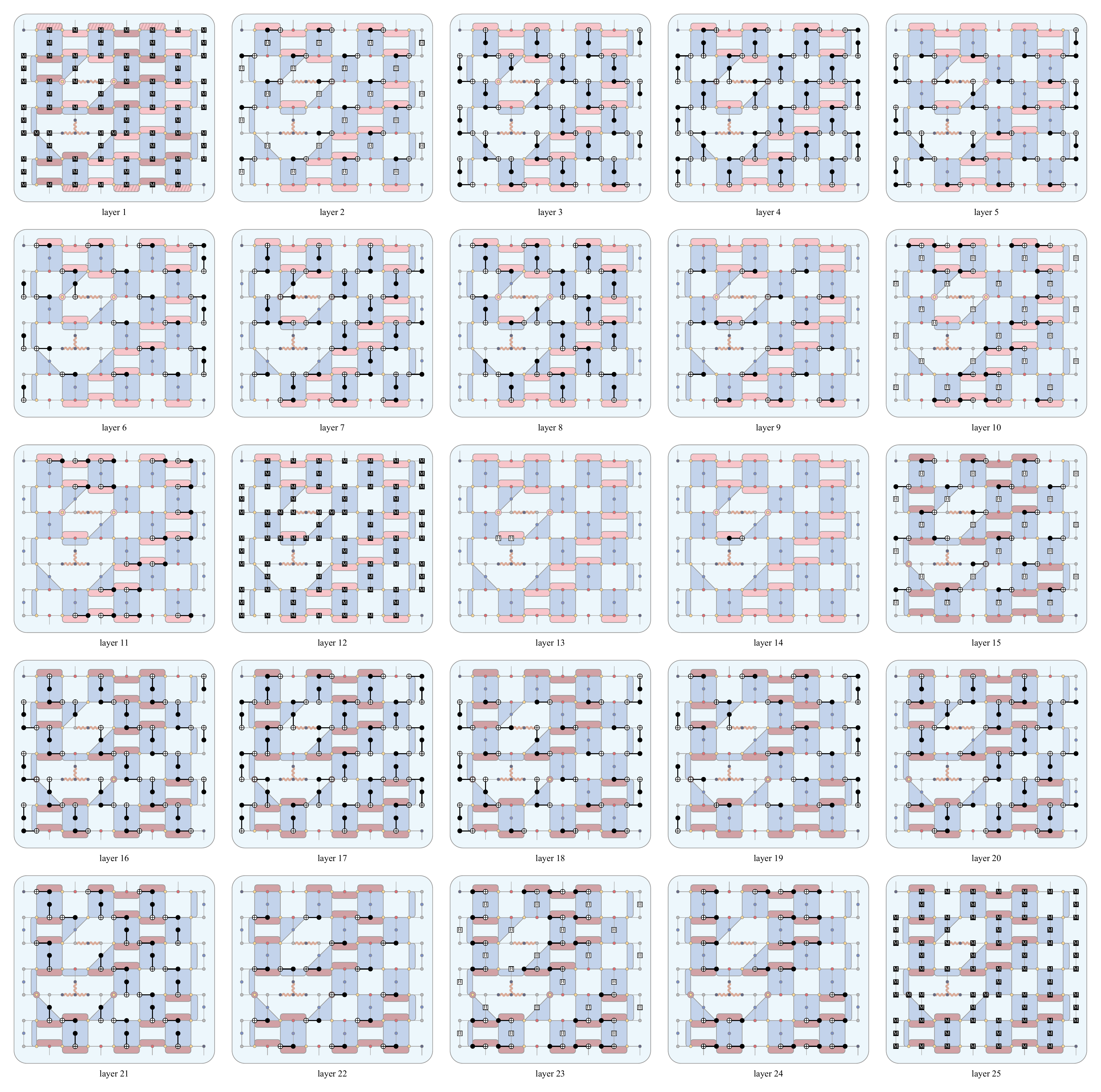}
    \caption{Syndrome measurement circuits used for $7\times 7$ memory experiments with defect exclusion (qubits 46 and 85 excluded). The circuit is overlayed on top of the checks measured and the device connectivity. Qubit color conventions match those of Fig.~\ref{fig:defect_exclusion_strat}, and we indicate defective couplers with red wavy lines. We show the full checks from Fig.~\ref{fig:defect_exclusion_strat}(c) in layer 1. In layers 2-14, the blue and light-red checks (shown) are measured (layers 13 and 14 are ``undoing" the Bell pair measurement; see Sec.~\ref{sec:bell_pair_mmt}). Layers 15-25 are measuring the blue and dark-red checks (shown).\label{fig:circuit_patch_overlay}}
\end{figure*}

In Equation~\ref{eqn:Bell_mmt_gadget}, we show the circuit used for Bell measurement. This circuit is chosen to help reduce overall gate depth of the entire syndrome measurement circuit. The final gate on the right-hand side can be combined with another gate in the full circuit (this gate arises from the measurement of the same two-body $X$-check in the next round; not shown) and both can be simply deleted. The Bell-measurement circuit works as intended if the middle qubit is in state $\ket{0}$. If it is in state $\ket{1}$, the $ZZ$ measurement outcome is flipped. Hence, we include the outcome of the middle qubit measurement in the $Z$ detector. Meanwhile, at the end of Equation~\ref{eqn:Bell_mmt_gadget}, the value of the $XX$ operator supported on the two data qubits is the product of the top two measurement outcomes. Hence the middle-qubit measurement outcome needs to be additionally included in subsequent $X$-detectors. The overall circuit when performing these Bell measurements is only increased in CNOT gate depth by 1 (see Fig.~\ref{fig:circuit_patch_overlay}).

\subsection{Lattice surgery mechanics}
There have been some different proposed implementations of lattice surgery in the literature e.g. \cite{bodeker2026, wang2026super}. 
Here we set out exactly the protocol we followed (see also Ref.~\onlinecite{zen2026low}). Unlike surface code lattice surgery, since we use the dynamic compass code (or the standard heavy-hex code), we can place two logical qubits in a vertical line without requiring an additional data-qubit strip between the two patches. 
This allows, for instance, two $3\times 3$ codes to be combined into a $3\times 6$ code during lattice surgery rather than a typical surface code setup which would require a $3\times 7$ combined code. \Cref{fig:lattice_surgery_loc} shows two of the placements we use. 

For the Bell fidelity experiments, we prepare each logical qubit in the code space by initializing the qubits in  $|0\rangle$ and performing an initial syndrome extraction round (which projects us into the code space).
We then perform one more full round of syndrome extraction which allows us to form a full set of both $X$-check (phase-flip) and $Z$-check (bit-flip) detectors.
The codes are then combined by measuring the stabilizers along the seam. 
After five syndrome extraction rounds of lattice surgery we stop measuring the checks across the seam, splitting the codes. 
We perform one more full extraction round on each of the logical qubits before measuring out the data qubits (in $Z$ or $X$ bases) and inferring the appropriate logical observable. If we measure out in the $Z$ basis, the observable is the product $\bar{Z}_1\bar{Z}_2$ obtained on readout. If we measure out in the $X$ basis, the observable is $\bar{X}_1\bar{X}_2$ from readout, combined with the outcome $m_X$ of that same logical operator obtained during lattice surgery. 
As described in Section~\ref{sec:bell_fidelity}, the Bell-state we have projected into is determined by the product of the central $X$-type merge-check outcomes on the seam. 
The reason we perform five rounds of syndrome extraction during the lattice surgery is because we are operating the device without resetting the measurement qubits. 
Five rounds are therefore required in order to maintain a minimum distance of 3.

\subsection{Bell Fidelity}\label{sec:bell_fidelity}

For each of the experiments we estimate the logical Bell-state correlators $\langle \bar{Z}_1\bar{Z}_2\rangle$ and $\langle \bar{X}_1\bar{X}_2\rangle$ from the corresponding readout circuits, after decoding. Our lattice surgery procedure is such that, in the absence of errors, $\langle \bar{Z}_1\bar{Z}_2\rangle = 1$ and $\langle \bar{X}_1\bar{X}_2\rangle = m_X = \pm 1$, depending on the outcome of the lattice surgery. Hence, we prepare the Bell state $\ket{\Phi^{m_X}} = \frac{1}{\sqrt{2}}(\ket{00} + m_X\ket{11})$.

For this target Bell state,
$$
  F_{\rm Bell}=\frac{1}{4}\bigl(1+\langle m_X\bar{X}_1\bar{X}_2\rangle+\langle \bar{Z}_1\bar{Z}_2\rangle
  -\langle m_X\bar{Y}_L\bar{Y}_L\rangle\bigr).
$$
A logical $Y$ measurement is not available fault-tolerantly in this setup \cite{lin2026surfacecodelogicaloperations}. Suppose the logical state is a mixed state written in the Bell basis, $\rho_L$. Writing $p_P$ as the probability associated with Bell basis state $(I\otimes P)|\Phi^{m_X}\rangle$ in $\rho_L$, it can be shown that $\langle \bar{X}_1\bar{X}_2\rangle=m_X(1-2(p_Y+p_Z))$, $\langle \bar{Z}_1\bar{Z}_2\rangle=1-2(p_X+p_Y)$, $\langle \bar{Y}_1 \bar{Y}_2\rangle = m_X(-1 + 2(p_X + p_Z))$ and therefore $F_{\rm Bell}=1-p_X-p_Y-p_Z$. We then obtain the following quantity from the measured logical expectation values:
$$
  F_{\rm lower}=\frac{\langle m_X \bar{X}_1\bar{X}_2\rangle+\langle \bar{Z}_1\bar{Z}_2\rangle}{2}
  = F_{\rm Bell}-p_Y .
$$

The residual gap is the logical-$Y$ error probability. From our above expressions for $\langle \bar{X}_1\bar{X}_2\rangle$ and $\langle \bar{Z}_1\bar{Z}_2\rangle$ we can bound $p_Y$ by $0\leq p_Y\leq \frac12(1-\max\{\langle m_X \bar{X}_L\bar{X}_L\rangle,
\langle \bar{Z}_L\bar{Z}_L\rangle\})$ (the upper bound is obtained from $p_Y \leq p_Y + p_X = \frac{1}{2}[1-\langle \bar{Z}_1\bar{Z}_2\rangle]$ and a similar expression involving $p_Z$). Hence, the measured Bell fidelity is rigorously bounded by $F_{\rm lower} \leq F_{\rm Bell} = F_{\rm lower} + p_Y \leq \frac12(1+\min\{\langle m_X\bar{X}_1\bar{X}_2\rangle, \langle \bar{Z}_1\bar{Z}_2\rangle)$.

\subsection{Pre-selection filters}\label{sec:prefilters}
Here we describe the various filters applied when performing pre-selection in lattice surgery experiments.

\subsubsection{Initial leakage filtering}

Leakage out of the computational subspace can limit the fidelities of QEC protocols by generating errors that are correlated over extended regions of space or across syndrome cycles. One way to sense for leakage is to check whether a $\pi$ pulse correctly flips the state of a qubit.

To improve the sensitivity of the protocol, we had access to an adiabatic, narrow-band $X$ gate, called the $X_{\rm {slow}}$ gate~\cite{barron2026observableestimationabsenceclassical}. Unlike a typical $X$ gate, which takes 32ns and has a spectral width of tens of MHz, the $X_{\rm{slow}}$ gate takes $\sim 1 \mu\text{s}$ and has a sub-MHz spectral width. The pre-check circuit then applies $X_{\rm{slow}}$-$X$-$M_{\rm{pre}}$ on each qubit. We wish to retain shots for which the measurement $M_{\rm{pre}}$ returns 0, and we wish to discard shots in any other condition, such as the presence of leakage, which results in measuring 1.

We note that measurement errors can also cause a qubit to fail this check. Therefore, we apply three repeated measurements in place of a single $M_{\rm{pre}}$. The filter is applied if at least two of the measurements return 1. In the Bell fidelity experiment, these pre-leakage filters removed approximately 50\% of the runs. Further details are reported in \cref{fig:lattice-heatmap}.

\subsubsection{Filtering on projection to the code-state}

We also pre-select on successful code-state preparation. Here we require that the initial detectors (both $Z$ and $X$) return trivial outcomes. Operationally, this removes shots with detected faults during the initial preparation and first syndrome-extraction rounds, before the lattice-surgery merge. In principle such a mechanism could be part of a prepare until succeed loop, carried out before the rest of the circuit is executed \cite{Gupta2024}. Alternatively, this procedure can be interpreted as an attempt to probe the logical failures resulting from lattice surgery, rather than poor initial state preparation. 

For the teleportation protocol, three different pre-filters were used: trivial syndrome in initial stabilizers in the prepared basis on the source logical qubit, trivial syndrome in initial stabilizers on the source \textit{and} target logical qubits, and trivial syndrome in the full set of initial detectors (that is, after the first 
two rounds of syndrome extraction, before lattice surgery is performed) on both logical qubits.

Details of the yields in each experiment resulting from these pre-filters can be found in Appendix~\ref{app:teleport}.

\begin{acknowledgements}
We are grateful for conversations with E. Pritchett, S. Singh and M. Takita.  BJB is grateful for the hospitality of the Niels Bohr Institute at the University of Copenhagen. We acknowledge support from the Intelligence Advanced Research Projects Activity (IARPA), under the Entangled Logical Qubits program through Cooperative Agreement Number W911NF-23-2-0223 (R.H., B.J.B., and S.D.B.).  The views and conclusions contained in this document are those of the authors and should not be interpreted as representing the official policies, either expressed or implied, of IARPA, the Army Research Office, or the U.S. Government. The U.S. Government is authorized to reproduce and distribute reprints for Government purposes notwithstanding any copyright notation herein. S.H.L was supported by the National Research Foundation of Korea (NRF) grant (RS-2026-25612364, RS-2026-25476454, RS-2024-00442710, RS-2023-NR068116) funded by the Ministry of Science and ICT (MSIT) of the Korean government.
\end{acknowledgements}

\appendix

\begin{figure}
    \centering
    \includegraphics[width=0.9\linewidth]{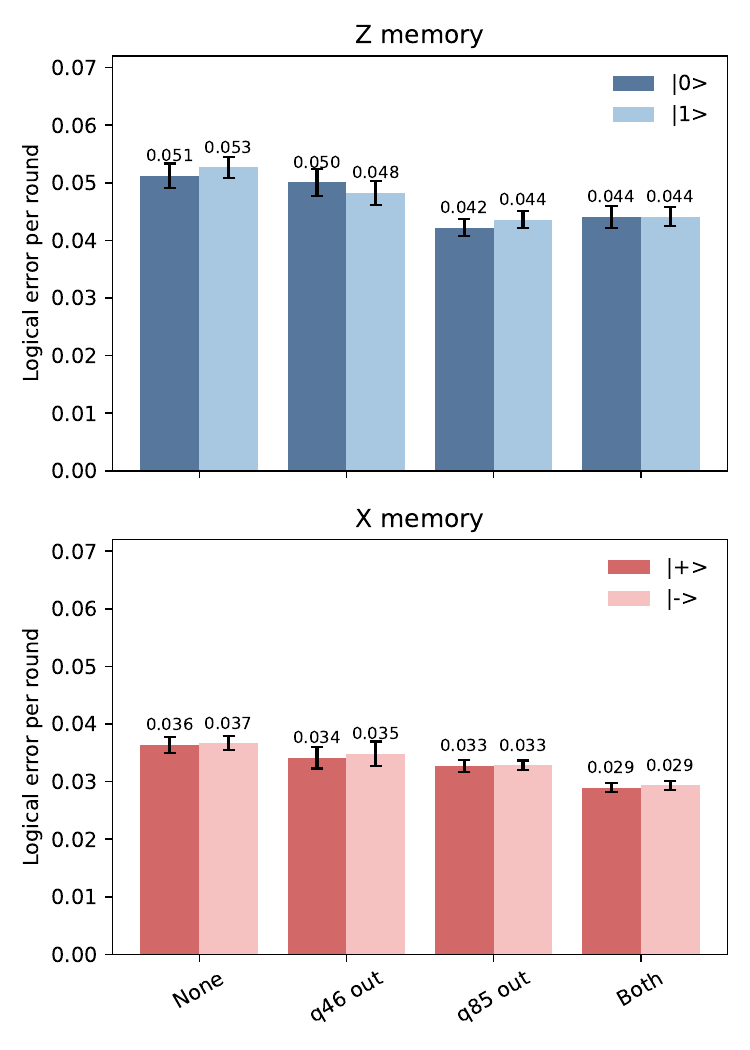}
    \caption{Comparison of defect exclusion strategies with data collected on 2026/05/01. We use a layer-averaged ACES noise model with ACES information obtained from a previous day's run, and a BeliefMatching decoder. The data show that the two basis states within a memory experiment differ very little.}
    \label{fig:may01_memory}
\end{figure}

\begin{figure}
    \centering
    \includegraphics[width=\linewidth]{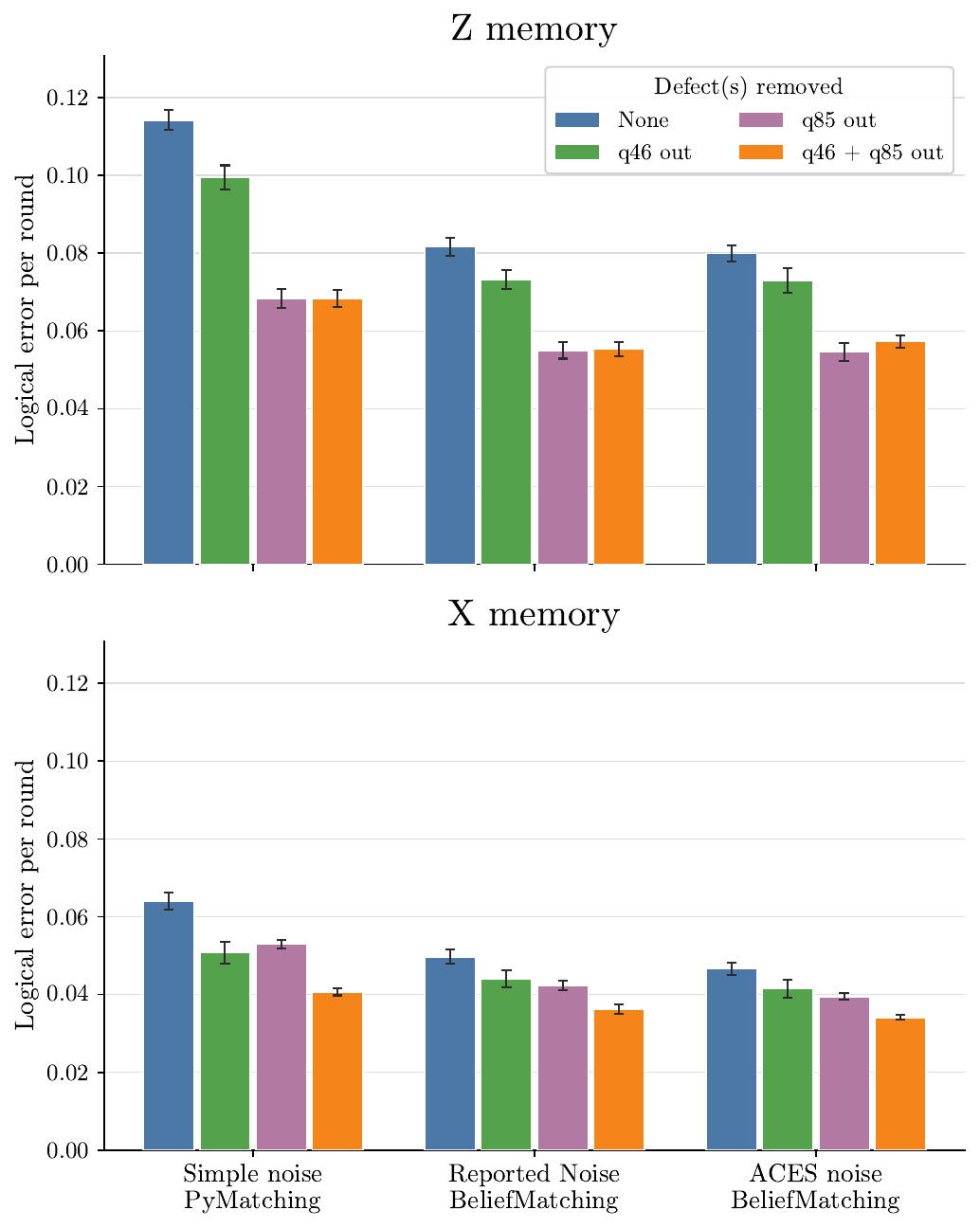}
    \caption{Further data comparing defect exclusion memory experiments and choices of decoder and noise model, collected on 2026/05/02. The decoders, noise models and defect exclusion circuit choices are explained in Section~\ref{sec:memory_results}.}
    \label{fig:may2_dropout_decoder_comparison_extra}
\end{figure}

\begin{figure}
    \centering
    \includegraphics[width=\linewidth]{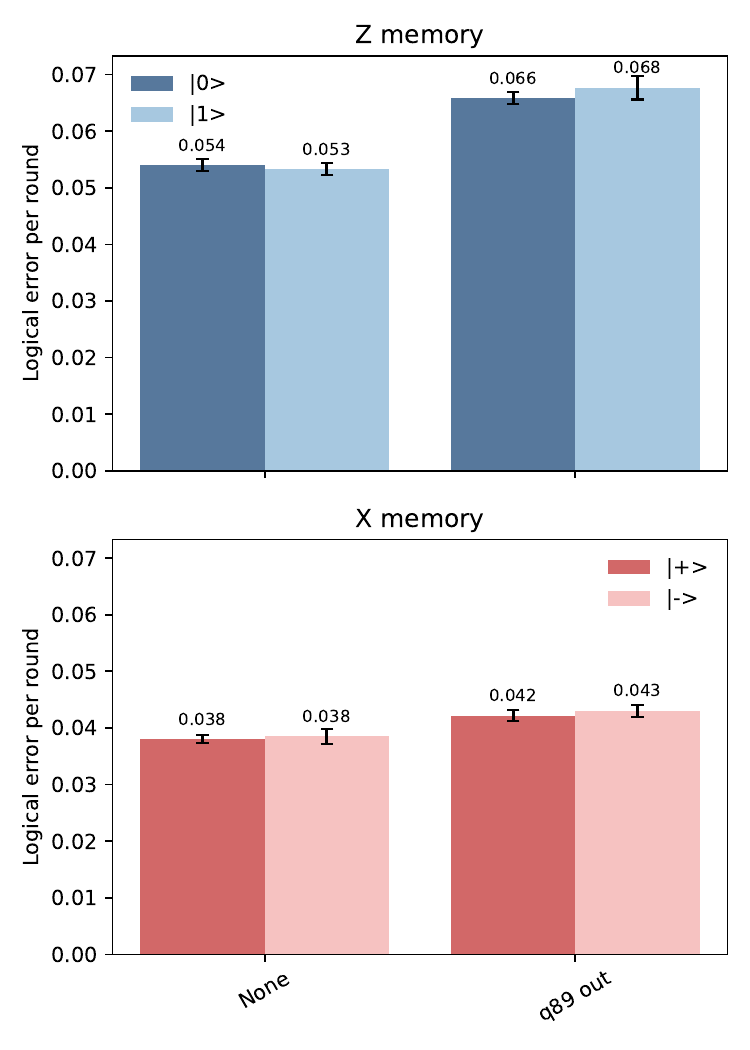}
    \caption{We compare the performance of the $7\times 7$ code with one with a ``dummy" defect excluded (qubit 89). We use a layer-averaged ACES noise model and a BeliefMatching decoder. Data collected on 2026/06/01.}
    \label{fig:Dummy_dropout}
\end{figure}

\begin{figure*}
    \includegraphics[width=0.95\textwidth]{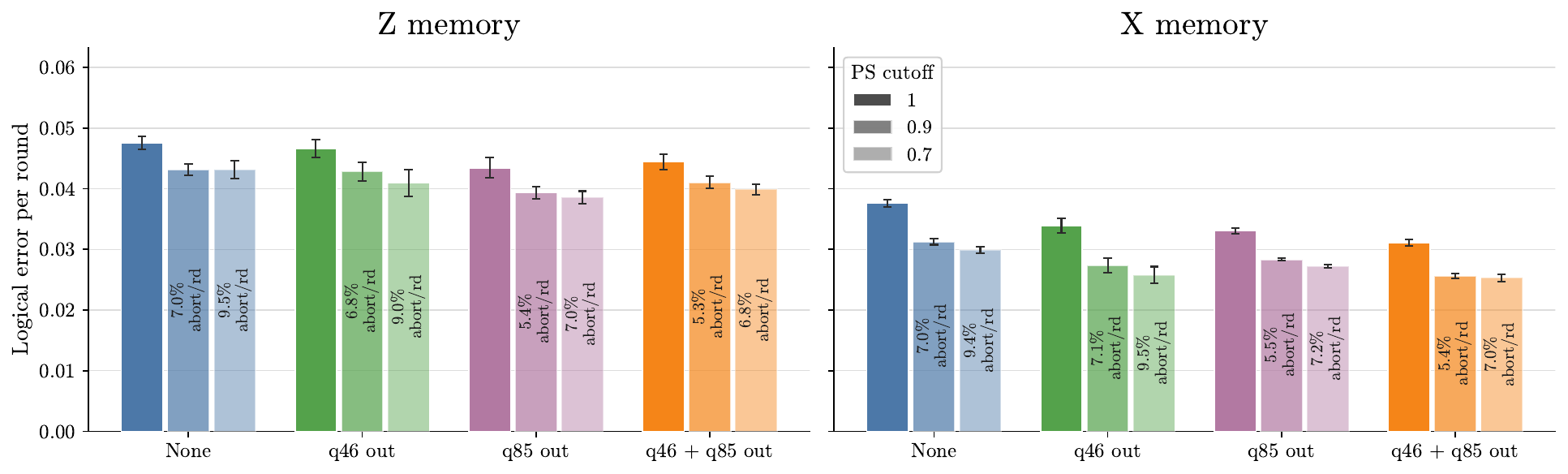}
    \caption{Leakage post-selection on patches with/without defect mitigation and in all bases. Data was taken on 2026/06/01.\label{fig:leakage_cutoff}}
\end{figure*}

\begin{figure*}
    \centering
    \begin{tikzpicture}
\node at (0,0) {\includegraphics[width=0.65\linewidth]{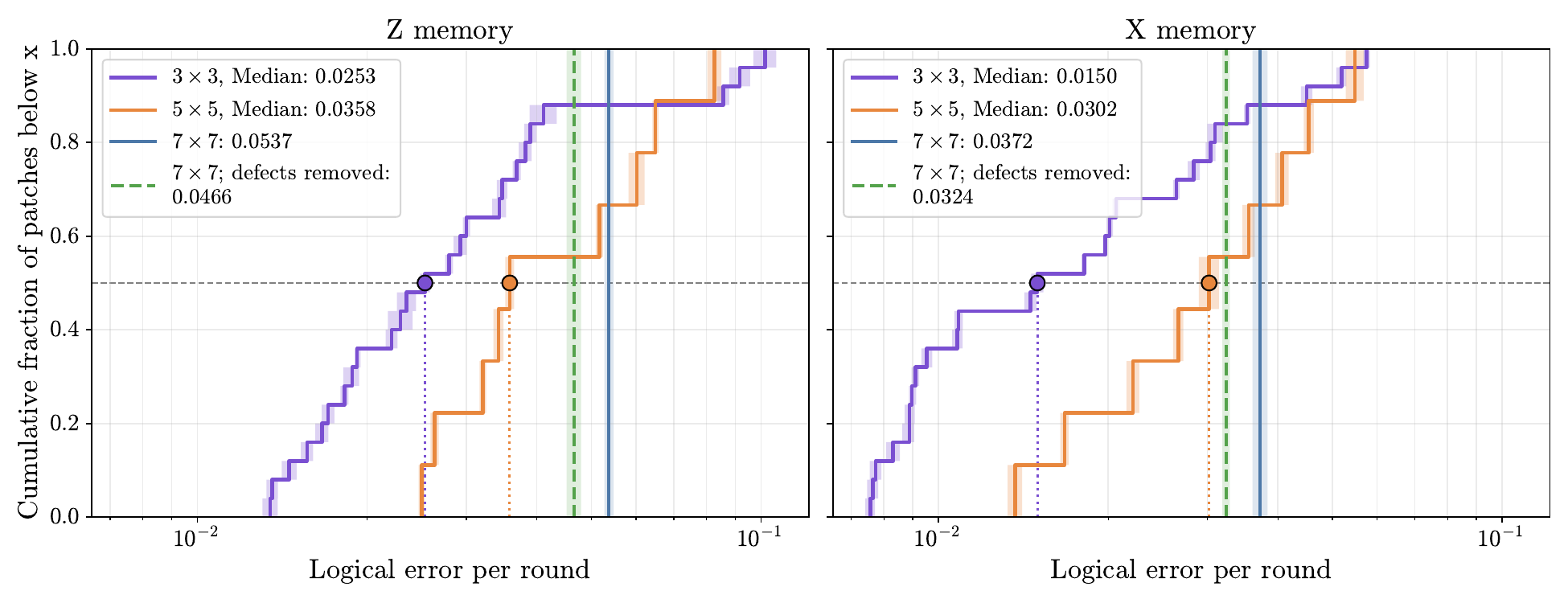}};
\node at (8.8,0) {\includegraphics[width=0.33\linewidth]{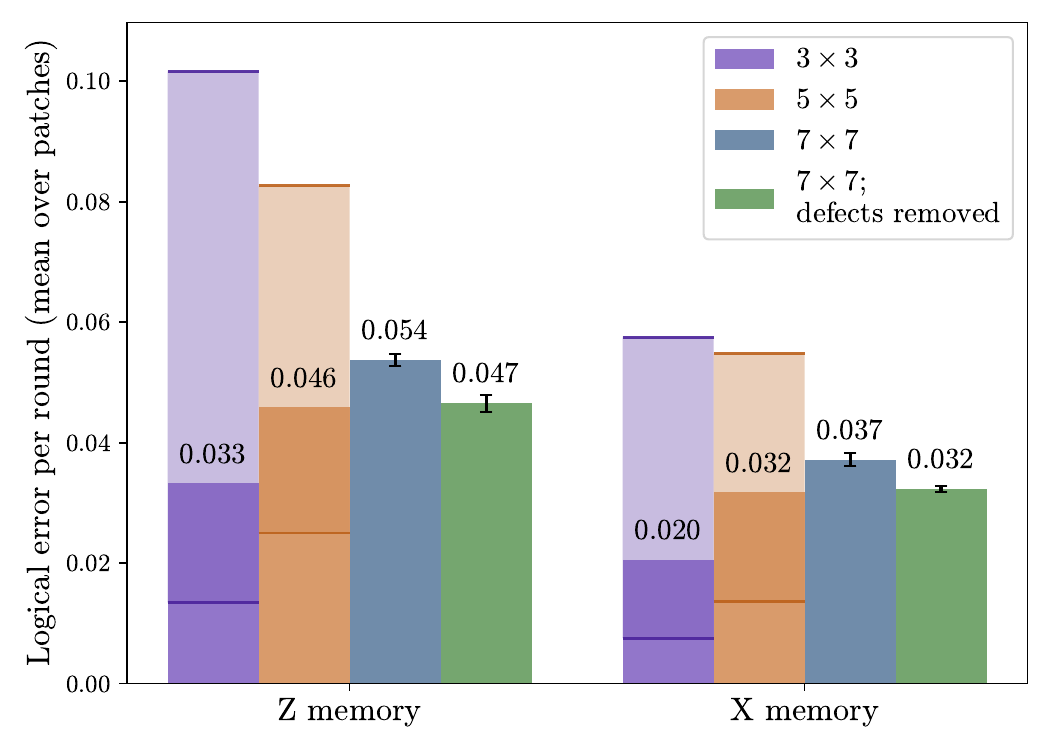}};
\node at (0,-2.5) {(a)};
\node at (8.5,-2.5) {(b)};
\end{tikzpicture}
    \caption{Comparison of dynamic compass code patches of side lengths 3, 5 and 7. The $7\times 7$ patch is compared both with and without defect exclusion. We report cumulative distributions of $3\times 3$ and $5\times 5$ patches in (a), and indicate the median patch performance with purple/orange circles. In (b), we plot mean patch performance (shaded regions indicate standard errors of the mean). Data was collected on 2026/06/01.}
    \label{fig:357_comparison}
\end{figure*}

\section{Further Memory Data}\label{app:further_mem_data}

\textit{Best memory run} --- Here, we present further data for memory experiments performed on \texttt{ibm\_boston}. In Fig.~\ref{fig:may01_memory}, we display results collected on 2026/05/01, a day which demonstrated the best performance for the $7\times 7$ code with both defects excluded. We find $4.4\%$ and $2.9\%$ logical errors per round for $Z$ and $X$ memory, respectively. The results were collected over 4096 shots. We did not perform separate experiments to extract an ACES noise model, and hence the presented results are decoded with an ACES noise model constructed on a previous day of device usage. From that old noise model, we perform averages over the layers in which each gate appears to construct a layer-averaged noise model we use for decoding. We do this to attempt to avoid overfitting the noise model to the previous day's device snapshot. We present the results for all four prepared basis states: $\ket{0}$, $\ket{1}$, $\ket{+}$ and $\ket{-}$.   

\textit{Best defect exclusion run} --- In Fig.~\ref{fig:may2_dropout_decoder_comparison_extra}, we provide further data used to generate Fig.~\ref{fig:dropout_vs_decoder}. We plot the logical error per round across four different defect exclusion circuits (excluding qubit 46, qubit 85, both or neither), and three different decoder+noise model combinations, as explained in Section~\ref{sec:memory_results}.

\textit{Dummy defect exclusion} --- We run control experiments in which we exclude non-defective components (around qubit 89) from the code in the same way as we exclude defects. We observe that the performance of this ``dummy" defect code is worse than the code with no qubit exclusion. This indicates that the improvements observed from defect exclusion are due to targeting noisy components of the lattice and not an artifact of moving to a lower-distance code in a device that is not operating below the code's threshold. We present the results from the dummy defect code in Fig.~\ref{fig:Dummy_dropout}.

\textit{Comparing leakage post-selection cutoffs} --- In Fig.~\ref{fig:leakage_cutoff}, we demonstrate that we observe significant gains from performing leakage post-selection with a cutoff probability of $0.9$, but observe little or no further gains from decreasing that cutoff to $0.7$. Meanwhile, the per-round abort probability continues to increase between cutoff $0.9$ and $0.7$.

\textit{$3\times 3$, $5\times 5$, and $7\times 7$ comparisons} --- Here we compare the memory performance of three different code sizes (both with and without defect mitigation). We provide the medians, means, and cumulative distributions of twenty-five $3\times 3$ and nine $5\times 5$ codes (within the footprint of the $7\times 7$ code) in Fig.~\ref{fig:357_comparison}. Note that data was collected on a different day to Fig.~\ref{fig:d5_d7_patches} of the main text.

\section{Further stability data}\label{app:more_stability_data}
\begin{figure*}
    \centering
    \includegraphics[width=0.9\linewidth]{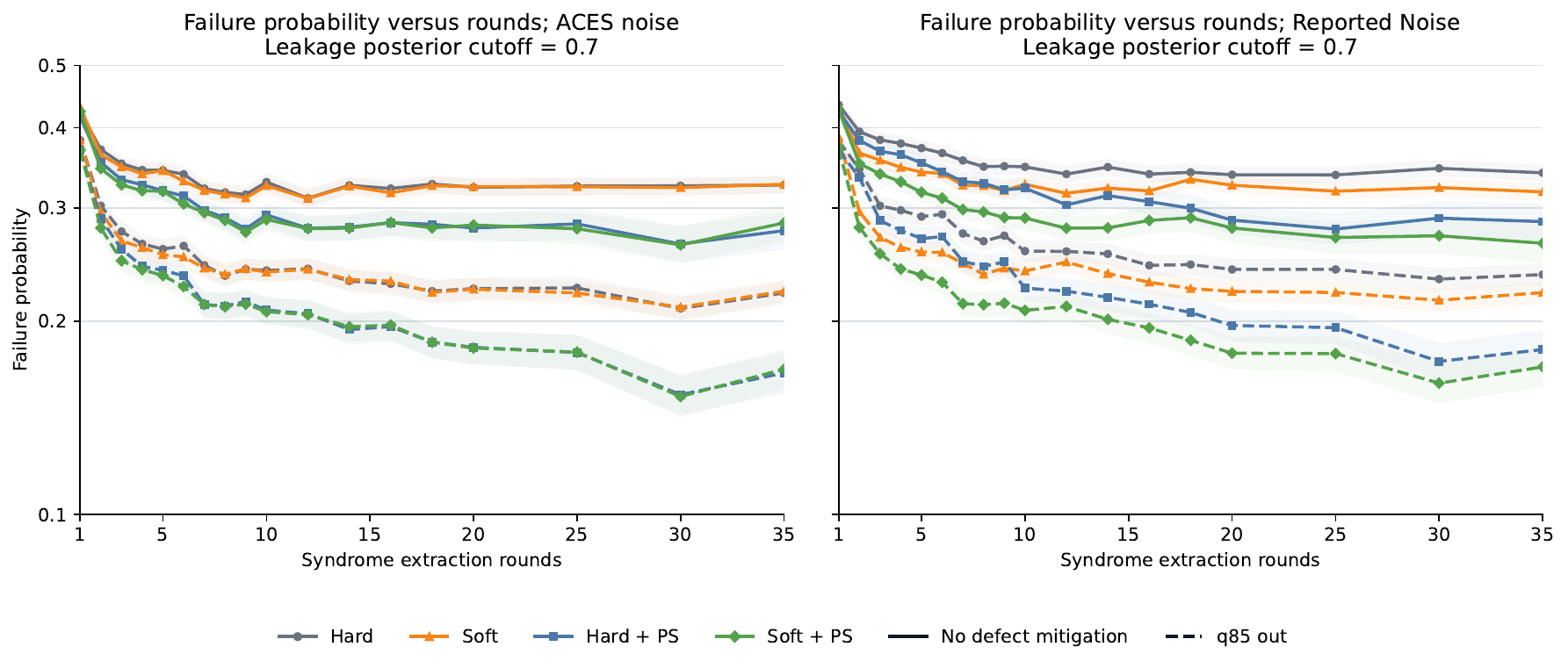}
    \caption{Further stability data. We compare decoders informed by ACES or reported noise, and hard or soft information. For each of these we decode the stability patch data with and without defect mitigation (qubit 85 excluded). We finally perform leakage post-selection (PS) on all sets of data with a cutoff probability of $0.7$. Data collected on 2026/06/01.}
    \label{fig:stability_more_data}
\end{figure*}

In Fig.~\ref{fig:stability_more_data}, we provide more data collected in our stability experiments. We compare decoders informed by both ACES noise and reported noise models. For each of these, we compare hard and soft decoding, with and without leakage post-selection (with a cutoff probability of $0.7$) and defect mitigation.

\section{Further Bell fidelity and logical teleportation data}\label{app:teleport}

\begin{figure*}
\begin{tikzpicture}
        \node at (0,6.0) {\includegraphics[width=0.95\textwidth]{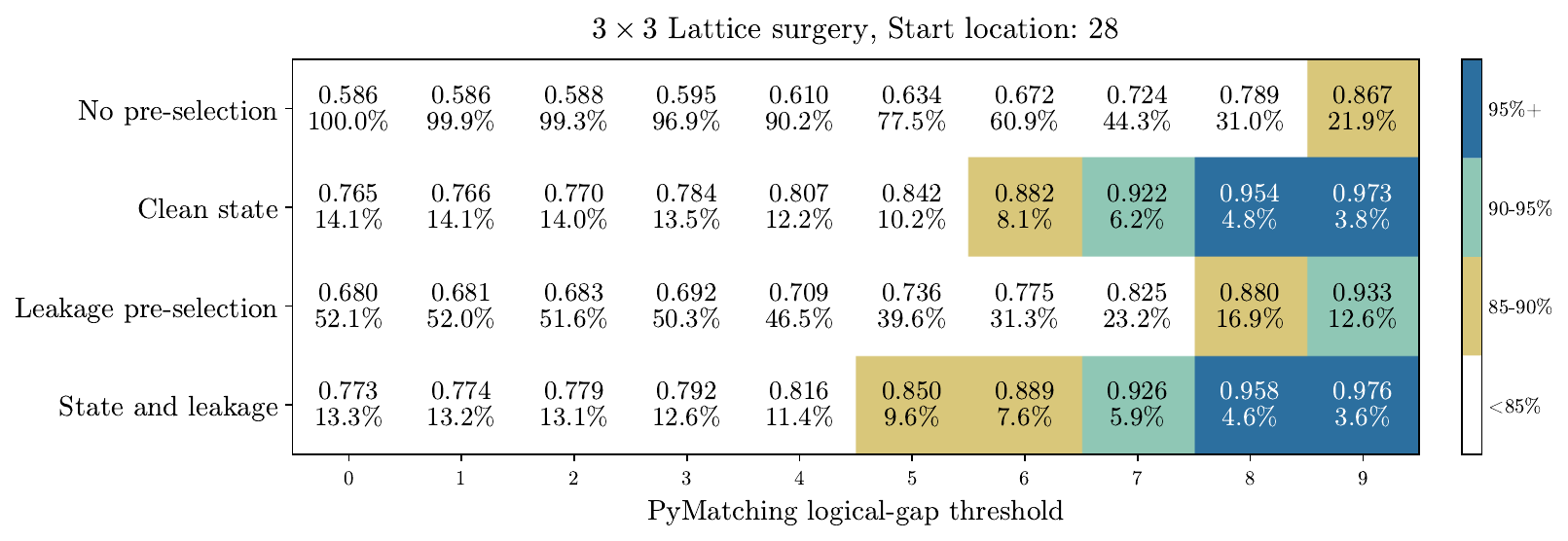}};
        \node at (0,0) {\includegraphics[width=0.95\textwidth]{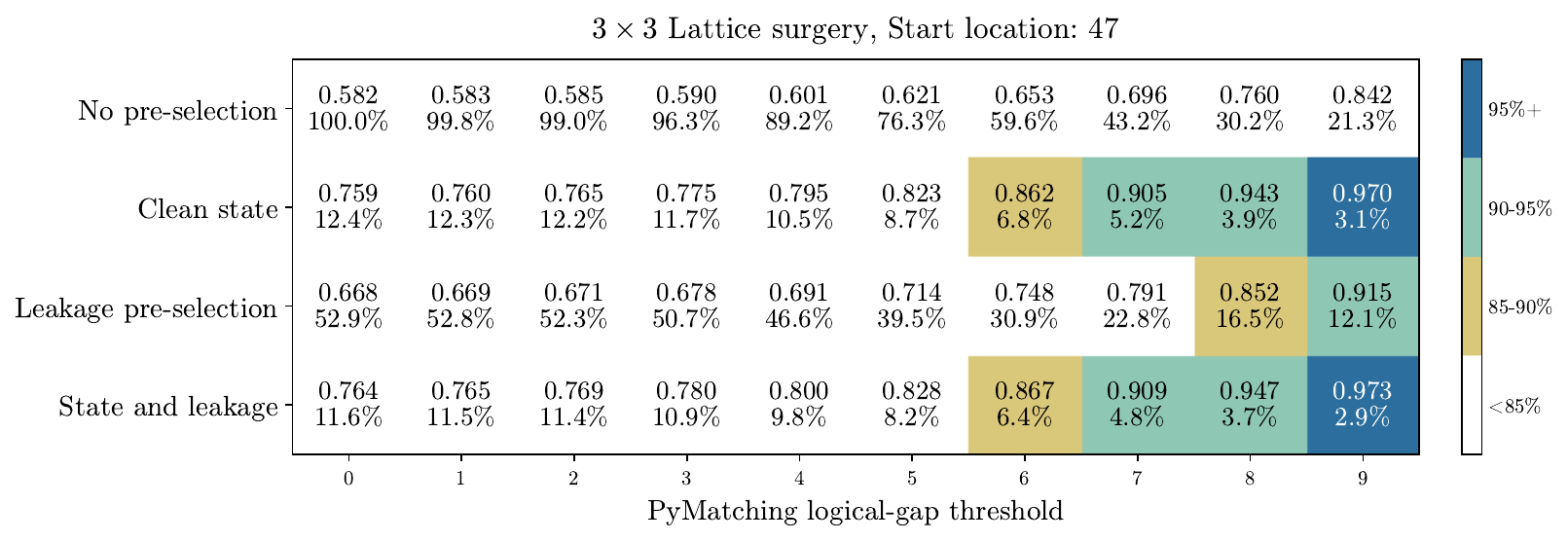}};
        
\end{tikzpicture}
\caption{Illustration of the impact of various pre-filters on the lower bound of the logical Bell Fidelity for two locations on the device (top-left qubit at location 28 or 47). The pre-filters are explained in Section~\ref{sec:prefilters} and include: applying no pre-selection filters, filtering on a clean preparation of the code space with no violated detectors (``clean state"), leakage pre-selection, and the combination of leakage and clean state pre-selection (``state and leakage"). As can be seen, most of the benefit comes from clean state pre-selection. 
Data was collected on 2026/07/17.}
\label{fig:lattice-heatmap}
\end{figure*}

\begin{figure*}
    \centering
    \begin{tikzpicture}
\node at (-4,0) {\includegraphics[width=0.6\linewidth]{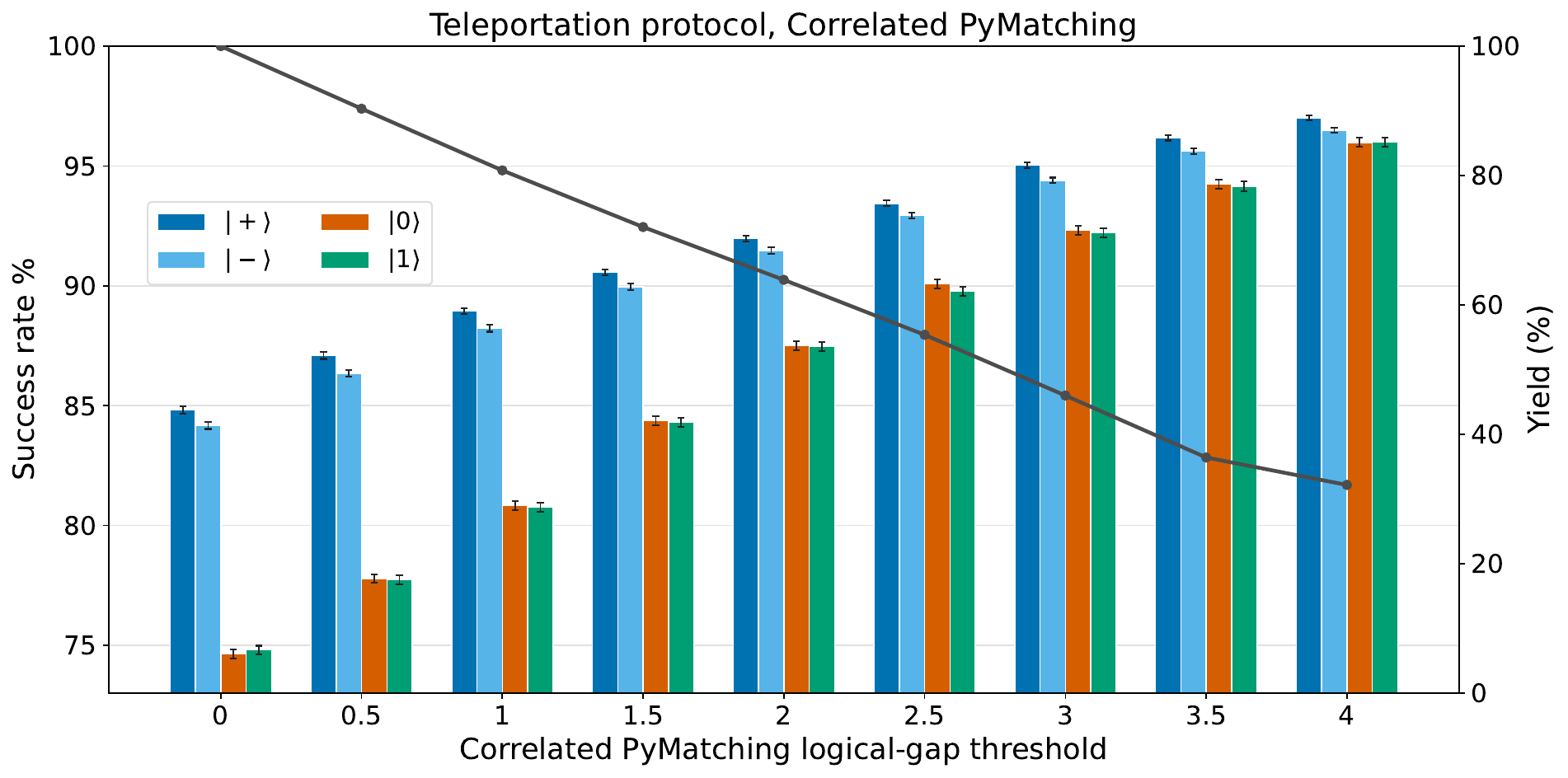}};
\node at (5,0) {\includegraphics[width=0.4\linewidth]{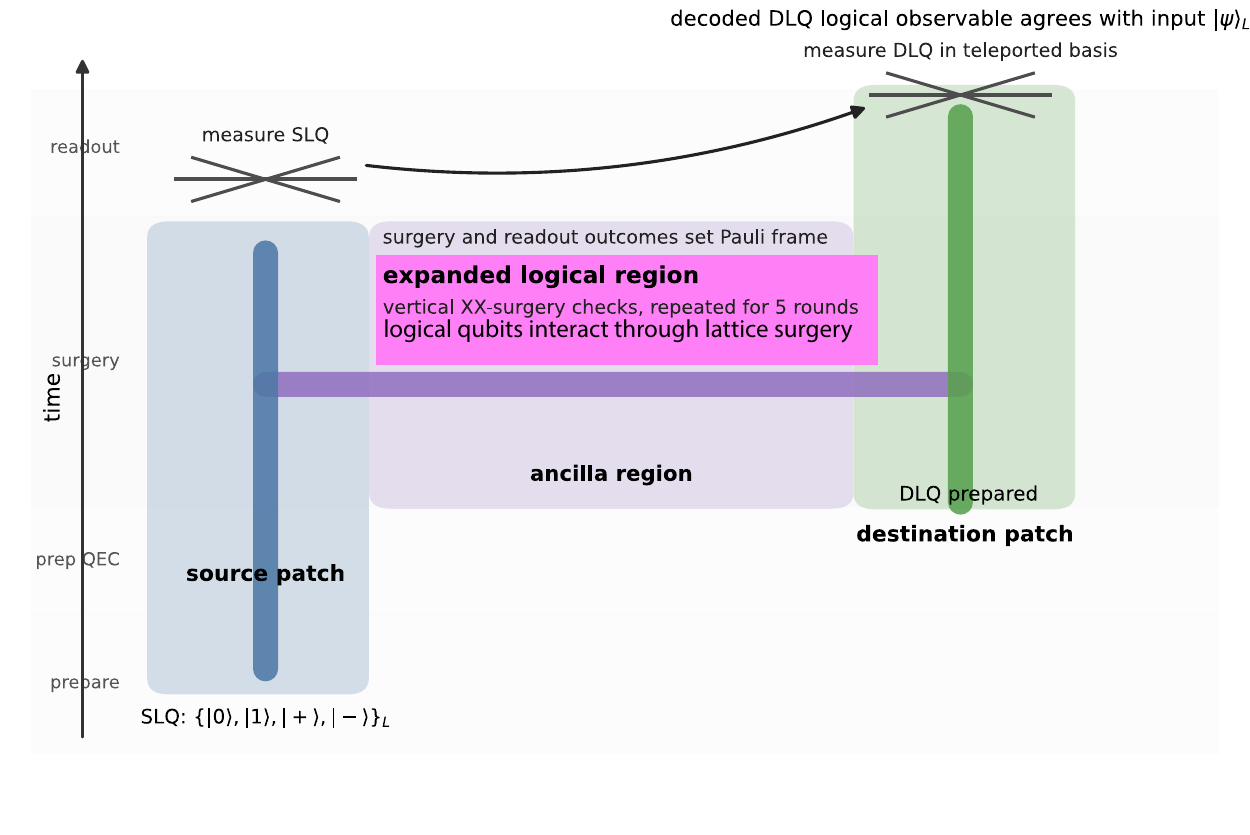}};
\node at (-1,-6.1) {\includegraphics[width=1\linewidth]{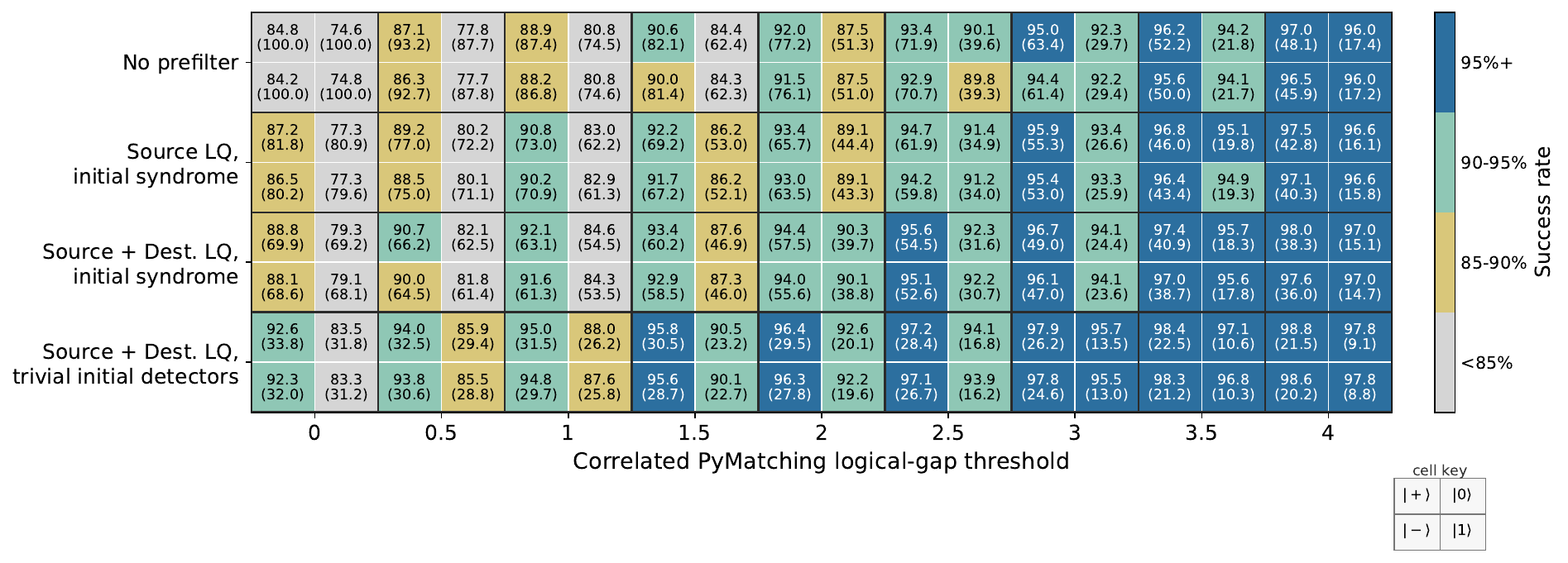}};
\node at (-9.5,2) {(a)};
\node at (1.5,2) {(b)};
\node at (-9.5,-3) {(c)};
\end{tikzpicture}
    \caption{Teleportation protocols success rates. Data was collected on 2026/07/22. (a) shows the success rate of teleporting various cardinal states. No pre-filter was applied. Correlated PyMatching was used to assess the logical gap and perform the decoding. The line plot indicates the yield at each threshold. (b) Cartoon pipe diagram showing the teleportation protocol. (c) Heat map of the teleportation success with various pre-filters applied (total yield in brackets), for each of the cardinal states. Each pre-filter, averaged over the four states, had the following yields: source logical qubit (LQ) initial syndrome 80.6\%; all qubits initial syndrome 68.9\%; all qubits initial syndrome plus the first full set of detectors 32.2\%.}
    \label{fig:teleport}
\end{figure*}

In Fig.~\ref{fig:lattice-heatmap}, we show the detailed results from pre- and post-selection filters for the $3\times 3$ lattice surgery experiments. We report results from two experiment locations: one with the top-left anchored at qubit 28 (shown in Fig.~\ref{fig:lattice_surgery_loc}), and another shifted down a row, with the top-left anchored at qubit 47. We note that pre-selecting on a clean preparation of the code state seemed to subsume the benefit from leakage pre-selection. For Bell fidelity experiments on $3\times 3$ codes, only approximately 14\% of the runs had trivial syndromes up to the first full set of detectors. Combining this filter with leakage pre-selection reduced this yield to 13\% with only a very mild additional improvement in the Bell fidelity (see Fig.~\ref{fig:lattice-heatmap}).

We also report results from logical state teleportation experiments in Fig.~\ref{fig:teleport}. Here we use lattice surgery (implementing an $\bar{X}_1\bar{X}_2$ measurement) between two logical qubits to teleport logical information from one logical qubit (source)  to the other (destination). For these experiments we only used the $3\times 3$ patches with all $Z$ checks measured every round (regular heavy-hex code).

We implement logical teleportation in the following way:
\begin{itemize}
    \item We prepare the source in one of the four cardinal states \(\{\ket{0}_L,\ket{1}_L,\ket{+}_L,\ket{-}_L\}\), and prepare the destination in the $\ket{0}_L$ state.
    \item The source is then coupled to the destination only through lattice surgery: the patches are expanded into a joint code region and stabilizers across the seam region are measured for five lattice surgery rounds.
    \item At the end of the surgery block we perform one further round of regular syndrome extraction on each logical qubit (with the two code patches now split).
    \item The data qubits belonging to the original source region are destructively measured in the $Z$ basis.
    \item Finally the destination logical qubit is measured in the same basis in which the source was prepared.
\end{itemize}
  
Performing lattice surgery for five rounds maintains the effective distance of the operation in the absence of mid-circuit reset.   See \cref{fig:teleport}(b) for a diagram showing the entire procedure.

A shot is counted as successful when the decoded destination logical observable agrees with the logical state initially prepared on the source.  
We repeat this procedure independently for the four input states and report the corresponding logical teleportation success probabilities.

We now comment on the pre- and post-selection filters and their yields for the logical teleportation protocol. In this experiment, we found that it was less important to pre-filter up to the first full set of detectors. Rather, the initial stabilizer readouts (in the same basis of state preparation) could be used for pre-filtering, with minimal further gain in fidelity found from pre-filtering further on a full set of stabilizers. Details of yields and the impact on fidelity of the three different pre-selection filters used for teleportation can be found in \cref{fig:teleport}(c) (see also Section~\ref{sec:prefilters}). One reason that the teleportation protocol is less sensitive to pre-selecting out errors in the opposite basis to Bell state preparation, is that the destination logical qubit is only ever read out in that basis. Contrast this with the Bell fidelity protocol where the logical qubits can be read out in the complimentary basis. 

For logical gap post-selection in teleportation experiments, PyMatching did not perform as well as correlated PyMatching in assessing the probability of corrections being accurate, so in this case we used correlated PyMatching~\cite{higgott2023} to calculate a logical gap. Shots that were accepted were then decoded also using correlated matching. In addition we found pre-selecting on correctly preparing the initial state was less important to achieve high fidelity teleportation. \Cref{fig:teleport}(a) shows the result of applying the logical gap post-selection without any other filters, and (c) shows the results of applying the filters together with logical gap post-selection.

In \Cref{fig:teleport}(c), we report the total yields from both pre-selection and post-selection. The yields from just post-selection can be calculated by comparing to the corresponding yields at logical-gap threshold 0. The average post-selection yields at threshold 4, for example, are: $32.2\%$ (no pre-filter), $35.6\%$ (source initial syndrome), $37.7\%$ (source and destination initial syndrome), and $45.9\%$ (both initial syndromes, plus trivial initial detectors).

\bibliography{hex}

\end{document}